\documentclass[runningheads]{llncs}

\usepackage[T1]{fontenc}
\usepackage{graphicx}
\usepackage{balance}
\usepackage{amsmath,amssymb,amsfonts}
\usepackage{textcomp}
\usepackage{booktabs}
\usepackage{multirow}
\usepackage{todonotes}
\usepackage{tabularx}
\usepackage{subfig}
\usepackage{graphicx}
\newcommand{\bm}[1]{\mathbf{#1}}
\usepackage{algorithmic}
\usepackage{color, colortbl}
\usepackage{sidecap}
\usepackage{rotating}
\usepackage[ruled,vlined,linesnumbered,noresetcount]{algorithm2e}
\SetKwInput{KwInput}{Input}                
\SetKwInput{KwOutput}{Output}              
\newsavebox\CBox
\def\textBF#1{\sbox\CBox{#1}\resizebox{\wd\CBox}{\ht\CBox}{\textbf{#1}}}
\usepackage[hypertexnames=false]{hyperref} 
\usepackage[capitalise,nameinlink,noabbrev]{cleveref}
\hypersetup{
	colorlinks=true,
	linkcolor=black,
	urlcolor=blue,
	citecolor=black
}
\usepackage[utf8]{inputenc}
\usepackage{scalefnt}
\usepackage{pgfplots}
\DeclareUnicodeCharacter{2212}{−}
\usepgfplotslibrary{groupplots,dateplot}
\usetikzlibrary{patterns,shapes.arrows}
\pgfplotsset{compat=newest}
\usepackage{lmodern}

\definecolor{LightOrange}{rgb}{1, 0.96, 0.87}

\newcolumntype{h}{>{\columncolor{LightOrange}}c}
\newcolumntype{k}{>{\columncolor{LightOrange}}l}

\newcommand{\subbest}[1]{\underline{\textBF{{#1}}}}
\newcommand{\unsubbest}[1]{\textBF{{#1}}}

\newcommand{\hlOAI}[1]{\textBF{{#1}}}
\newcommand{\hlCXR}[1]{\underline{{#1}}}

\usepackage{listings}
\begin{document}
	\Crefname{figure}{Suppl. Figure}{Suppl. Figures}
	\Crefname{table}{Suppl. Table}{Suppl. Tables}
	\Crefname{section}{Suppl. Section}{Suppl. Sections}
	\Crefname{proposition}{Suppl. Proposition}{Suppl. Propositions}
	\Crefname{algorithm}{Suppl. Algorithm}{Suppl. Algorithms}
	
	\title{AdaTriplet: Adaptive Gradient Triplet Loss \\ with Automatic Margin Learning for Forensic Medical Image Matching}
	\titlerunning{AdaTriplet: Adaptive Gradient Triplet Loss}
	
	
	%
	\author{Khanh Nguyen\thanks{Equal contributions} \and
		Huy Hoang Nguyen$^\star$\and
		Aleksei Tiulpin}
	\authorrunning{Nguyen et al.}
	\institute{University of Oulu, Oulu, Finland\\
		\email{\{khanh.nguyen,huy.nguyen,aleksei.tiulpin\}@oulu.fi}}
	\maketitle              
	\begin{abstract}
		This paper tackles the challenge of forensic medical image matching (FMIM) using deep neural networks (DNNs). FMIM is a particular case of content-based image retrieval (CBIR). The main challenge in FMIM compared to the general case of CBIR, is that the subject to whom a query image belongs may be affected by aging and progressive degenerative disorders, making it difficult to match data on a subject level. CBIR with DNNs is generally solved by minimizing a ranking loss, such as Triplet loss (TL), computed on image representations extracted by a DNN from the original data. TL, in particular, operates on triplets: anchor, positive (similar to anchor) and negative (dissimilar to anchor). Although TL has been shown to perform well in many CBIR tasks, it still has limitations, which we identify and analyze in this work. In this paper, we introduce (i) the AdaTriplet loss -- an extension of TL whose gradients adapt to different difficulty levels of negative samples, and (ii) the AutoMargin method -- a technique to adjust hyperparameters of margin-based losses such as TL and our proposed loss dynamically. Our results are evaluated on two large-scale benchmarks for FMIM based on the Osteoarthritis Initiative and Chest X-ray-14 datasets. The codes allowing replication of this study have been made publicly available at \url{https://github.com/Oulu-IMEDS/AdaTriplet}.
	\end{abstract}
	
	\section{Introduction}
	
	Content-based image retrieval (CBIR) describes the long-standing problem of retrieving semantically similar images from a database. CBIR is challenging due to the diversity of foreground and background color, context, and semantic changes in images~\cite{schroff2015facenet}. Besides general computer vision~\cite{saritha2019content,tzelepi2018deep}, in the domain of medicine content-based medical image retrieval (CBMIR)is growing~\cite{choe2022content,zhang2022content}, due to the increasing demand for effectively querying medical images from hospital picture archive and communication systems (PACS)~\cite{hostetter2018integration}. 
	
	In CBMIR, given a medical image (query), one aims to search in a database for images that are similar disease-wise or belonging to the same subject. The former problem is related to diagnostic applications, and the latter problem is of interest for forensic investigations. Hereinafter, we name this problem forensic medical image matching (FMIM). Unlike general CBIR, longitudinal medical imaging data of a person evolves in time due to aging and the progression of various diseases (see~\cref{fig:matched_samples}). Therefore, the FMIM domain poses new challenges for CBIR.
	
	Deep learning (DL)-based methods have made breakthroughs in various fields, and in particular metric learning, which is the backbone of CBIR~\cite{choe2022content,liang2021exploring,tzelepi2018deep,zhang2022content}. The aim of DL-based metric learning is to train a functional parametric mapping $f_\theta$ from the image space $\mathbb{R}^{C\times H \times W}$ to a lower-dimensional feature space $\mathbb{R}^D$. In this feature space, representations of semantically similar images are close, and ones of irrelevant images are distant. In our notation, $C$, $H$ and $W$ represent the number of channels, height, and width of an image, respectively.
	
	\newdimen\figrasterwd
\figrasterwd\textwidth

\begin{figure}[t!]
    \centering
    \IfFileExists{figures/gradient_fields/Triplet_m25-crop.pdf}{}{\immediate\write18{pdfcrop 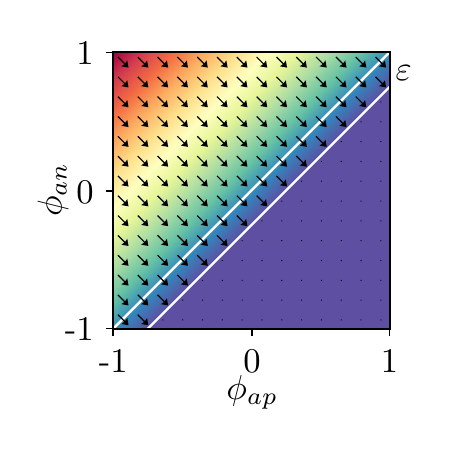}}
    \IfFileExists{figures/gradient_fields/Lan_m10-crop.pdf}{}{\immediate\write18{pdfcrop 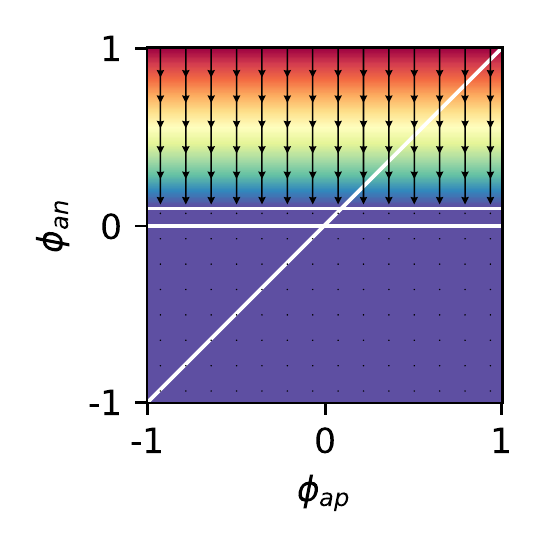}}
    \IfFileExists{figures/gradient_fields/Triplet_Lan_m25-crop.pdf}{}{\immediate\write18{pdfcrop 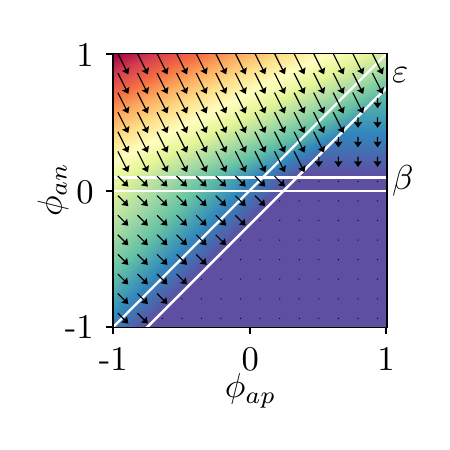}}
    \IfFileExists{figures/samples_cxr/CXR_36_7124_96_PA_short-crop.pdf}{}{\immediate\write18{pdfcrop 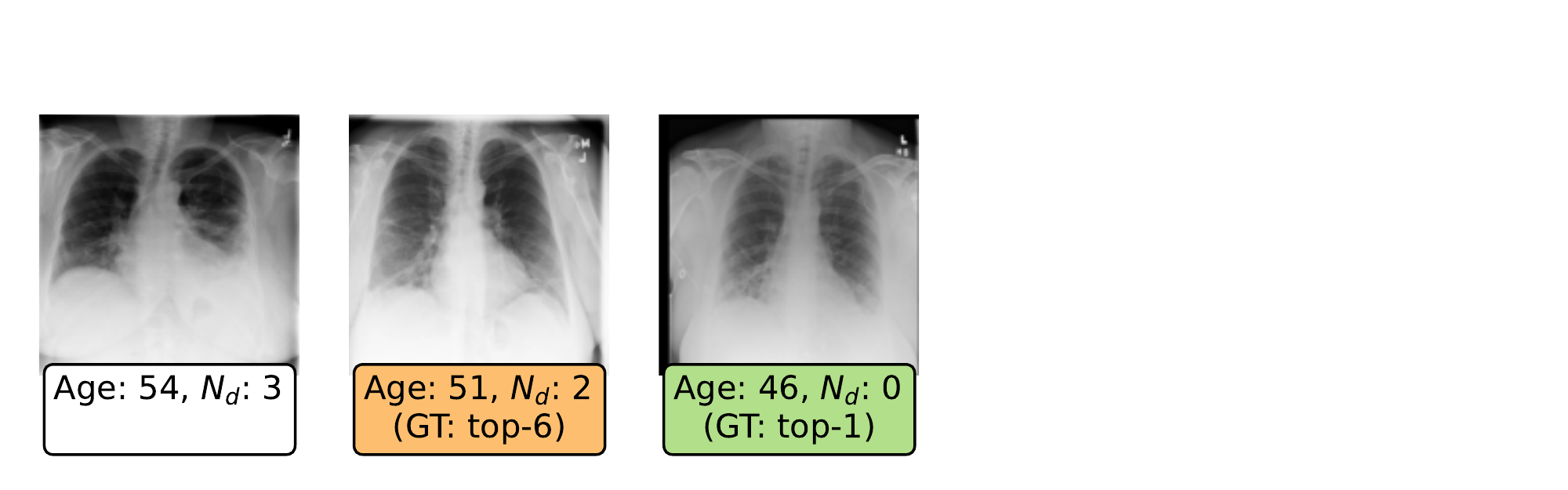}}
    \IfFileExists{figures/samples_cxr/CXR_162_1946_108_PA_short-crop.pdf}{}{\immediate\write18{pdfcrop 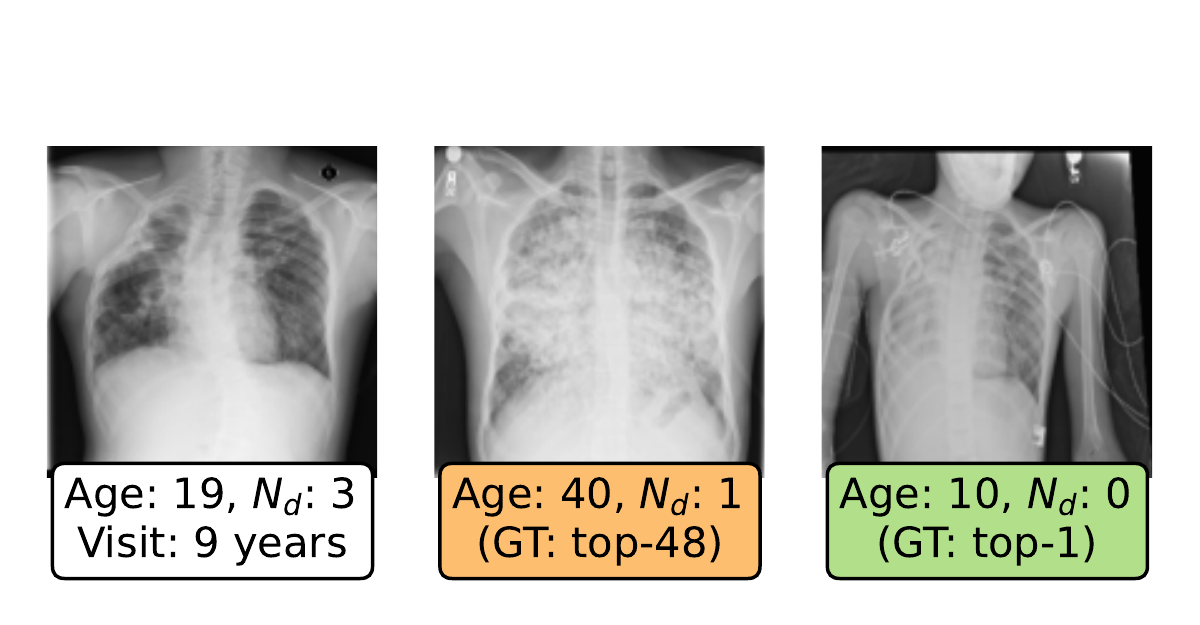}}
    \IfFileExists{figures/samples_oai/OAI_44_9271853_96_R_short-crop.pdf}{}{\immediate\write18{pdfcrop 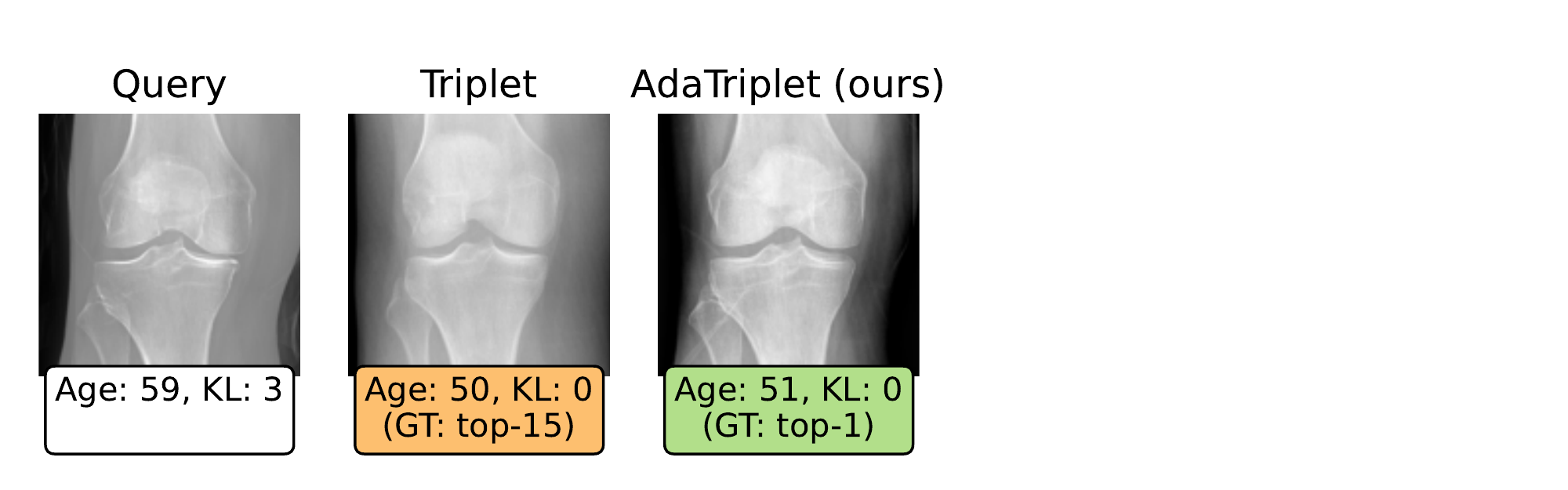}}
    \IfFileExists{figures/samples_oai/OAI_47_9378288_96_R_short-crop.pdf}{}{\immediate\write18{pdfcrop 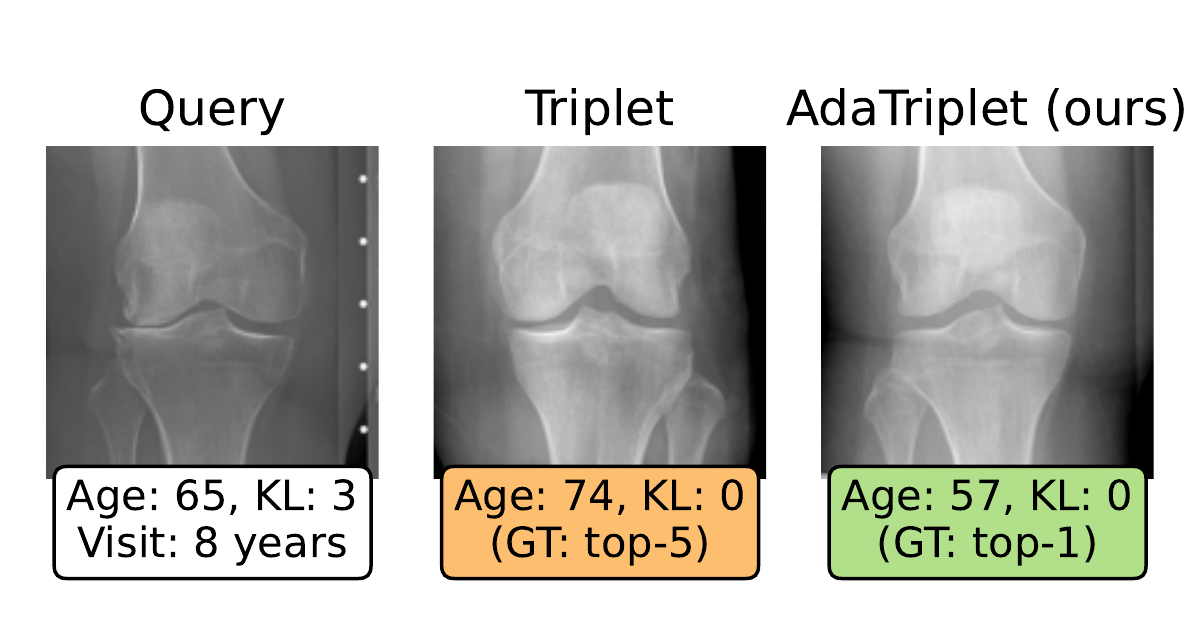}}
    
        \parbox{\figrasterwd}{
        \parbox{.35\figrasterwd}{%
      \includegraphics[width=.35\textwidth]{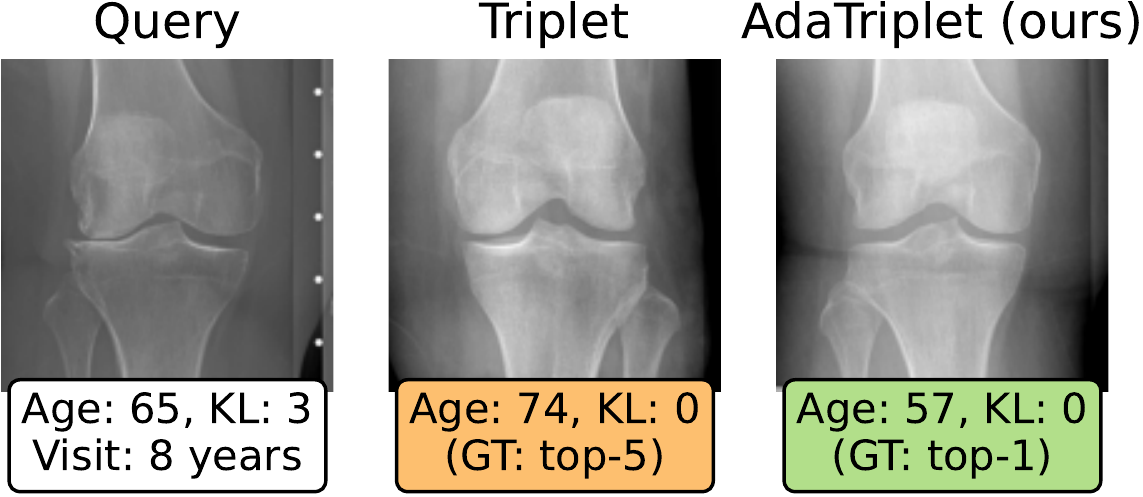}\\
      \subfloat[\small FMIM samples\label{fig:matched_samples}]{\includegraphics[width=.34\textwidth]{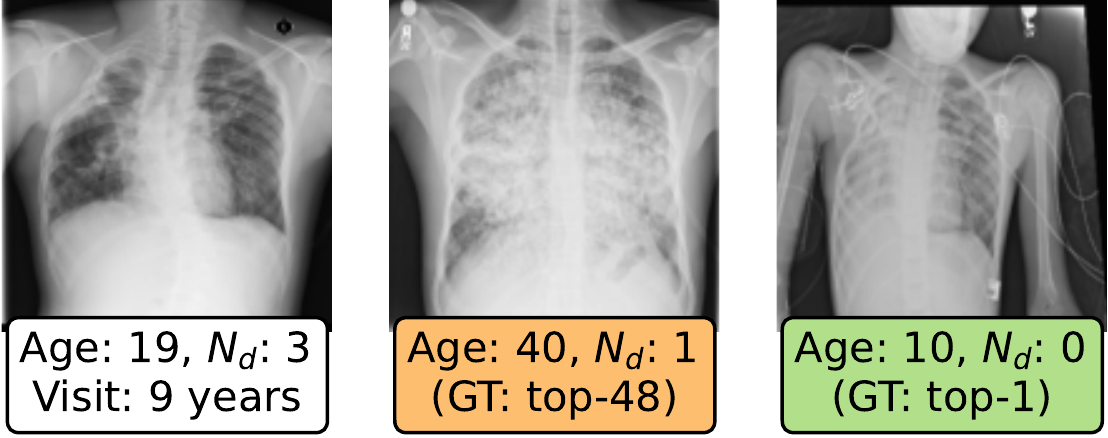}}
    }
    \hfill
        \parbox{.3\figrasterwd}{%
        \subfloat[\small Triplet loss~\eqref{eq:triplet_loss_norm}\label{fig:gradient_field_triplet}]{\includegraphics[width=\hsize, height=\hsize]{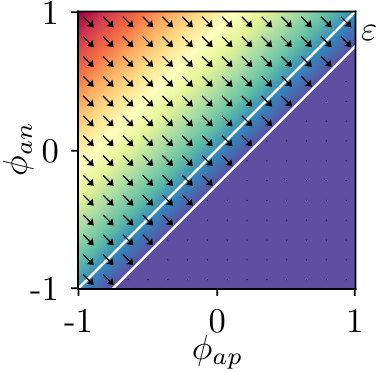}}
        }
        \parbox{.3\figrasterwd}{%
          \subfloat[\small AdaTriplet loss~\eqref{eq:our_triplet_loss_norm}\label{fig:gradient_field_ours}]{\includegraphics[width=\hsize, height=\hsize]{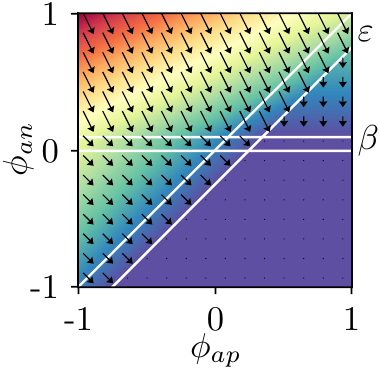}}
        }
    }
    \caption{\small Comparisons between the Triplet loss and our AdaTriplet loss. (a) Top-1 retrieved results. Green: if a ground truth (GT) is the top-1 in the ranked retrieval list, orange: otherwise. KL indicates the grade of knee osteoarthritis severity. $N_d$ is the number of thorax diseases. (b-c) 2D loss surfaces and negative gradient fields of the two losses. 
    Each point is a triplet. 
    Loss values are represented by colors (increasing from purple to red). The arrows are negative gradient vectors. }
    \label{fig:gradient_fields}
    \vspace{-3mm}
\end{figure}
	
	The loss function is the central component of metric learning~\cite{musgrave2020metric}, and there exist two major types: (i) those that enforce the relationships between samples in each batch of data during stochastic optimization -- \emph{embedding losses}~\cite{hoffer2015deep,liang2021exploring,schroff2015facenet,xuan2020hard,zhang2022content} and (ii) \emph{classification losses}~\cite{deng2019arcface,tzelepi2018deep,zhao2019weakly}. Two fundamental embedding losses that previous studies have built upon are Contrastive loss (CL)~\cite{chopra2005learning} and Triplet loss (TL)~\cite{hoffer2015deep}. The idea of the CL, is to minimize the feature space distance between similar data points, and maximize it for the dissimilar ones. The TL, on the other hand, considers every triplet of samples -- anchor, positive and negative, and aims to ensure that the distance between the anchor and positive samples is smaller than the distance between the anchor and negative ones. 
	
	In many practical applications, although the TL is more commonly used than the CL~\cite{bai2020deep,yuan2020defense}, it also has limitations. Firstly, the TL depends on a ``margin'' hyperparameter, which is usually fixed and needs to be chosen empirically. Secondly, as we show in this work, the TL ignores the magnitude of the pair-wise distances, thus may overlook the case where anchors and negative samples are too close. In this paper, we tackle these limitations, and summarize our contributions as follows:
	\begin{enumerate}
		\item We theoretically analyze the TL, and propose \emph{an adaptive gradient triplet loss}, called AdaTriplet, which has appropriate gradients for triplets with different hardness. That characteristic makes our loss distinct from the TL, as illustrated in~\cref{fig:gradient_fields}.
		\item To address the issue of selecting margin hyperparameters, we propose a simple procedure -- AutoMargin, which estimates margins adaptively during the training process, and eliminates the need for a separate grid-search.
		\item Through a rigorous experimental evaluation on knee and chest X-ray image forensic matching problems, we show that AdaTriplet and AutoMargin allow for more accurate FMIM than a set of competitive baselines.
	\end{enumerate}

	\section{Methods}
	\subsection{Problem Statement}
	Let $\bm{X}\times \bm{Y} = \{(\bm{x}_i, y_i)\}_{i=1}^{N}$ be a dataset of medical images $\bm{x}_i$'s $\in \mathbb{R}^{C\times H\times W}$ and corresponding subjects' identifiers $y_i$'s with $|\bm{Y}|\leq N$. We aim to learn a parametric mapping $f_\theta : \mathbb{R}^{C\times H\times W} \xrightarrow{} \mathbb{R}^{D}$ such that $\forall (\bm{x}_i, y_i), (\bm{x}'_i, y_i), (\bm{x}_j, y_j) \in \bm{X}_{train} \times \bm{Y}_{train}, \bm{X}_{train} \subset \bm{X}, y_i \neq y_j$,
	\begin{align}\label{eq:triplet_ineq}
	d(f_\theta(\bm{x}_i) , f_\theta(\bm{x}'_i)) < d(f_\theta(\bm{x}_i) , f_\theta(\bm{x}_j)).
	\end{align}
	
	The learned mapping $f_\theta$ is expected to be generalizable to $\bm{X}_{test} = \bm{X} \setminus \bm{X}_{train}$ where $\bm{Y}_{test} \cap \bm{Y}_{train} = \varnothing$. Often, $\bm{x}_i$ is called an anchor point, $\bm{x}'_i$ -- a positive point, and $\bm{x}_j$ -- a negative point. Hereinafter, they are denoted as $\bm{x}_a, \bm{x}_p$, and $\bm{x}_n$, respectively. For simplicity, we also denote $\bm{f}_a = f_\theta(\bm{x}_a)$, $\bm{f}_p = f_\theta(\bm{x}_p)$, $\bm{f}_n = f_\theta(\bm{x}_n)$, $\phi_{ap} = \bm{f}_a^\intercal\bm{f}_p$, and $\phi_{an} = \bm{f}_a^\intercal\bm{f}_n$.
	
	\subsection{Triplet Loss}
	Let $\mathcal{T}=\left \{(\bm{x}_a, \bm{x}_p, \bm{x}_n) \mid y_a = y_p, y_a \neq y_n \right \}$ denote a set of all triplets of an anchor, a positive, and a negative data point. For each $(\bm{x}_a, \bm{x}_p, \bm{x}_n) \in \mathcal{T}$, the Triplet loss is formulated as~\cite{chechik2010large,hoffer2015deep}:
	\begin{align}
	\mathcal{L}_{\mathrm{Triplet}} = \left [ \|\bm{f}_a - \bm{f}_p \|_2^2 - \| \bm{f}_a - \bm{f}_n \|_2^2 + \varepsilon \right ]_{+},
	\label{eq:triplet_loss}
	\end{align}
	where $\left [ \cdot \right ]_{+} = \max(\cdot , 0)$, and $\varepsilon$ is a non-negative margin variable. Following common practice, we normalize all feature vectors, that is $\left \| \bm{f}_a\right \|_2 = \left \| \bm{f}_p\right \|_2 = \left \| \bm{f}_n\right \|_2 = 1$, as well since we can then derive that $\varepsilon \in [0,4)$. Thereby, we can convert Eq.~\eqref{eq:triplet_loss} to a slightly different objective, which is identical to optimize, but allows us to identify limitations of the TL. 
	\begin{proposition}
		Given $\left \| \bm{f}_a\right \|_2 = \left \| \bm{f}_p\right \|_2 = \left \| \bm{f}_n\right \|_2 = 1$, minimization of the Triplet loss \eqref{eq:triplet_loss} corresponds to minimizing
		\begin{align}
		\mathcal{L}_{\mathrm{Triplet}}^* = \left [\phi_{an} - \phi_{ap}+ \varepsilon \right ]_{+},\ \varepsilon \in [0, 2).
		\label{eq:triplet_loss_norm}
		\end{align}
		\begin{proof} See~\Cref{proof:prop1}.
		\end{proof}
	\end{proposition}
	
	\begin{figure}[t!]
		\centering
		\IfFileExists{figures/automargin/distribution_delta-crop.pdf}{}{\immediate\write18{pdfcrop 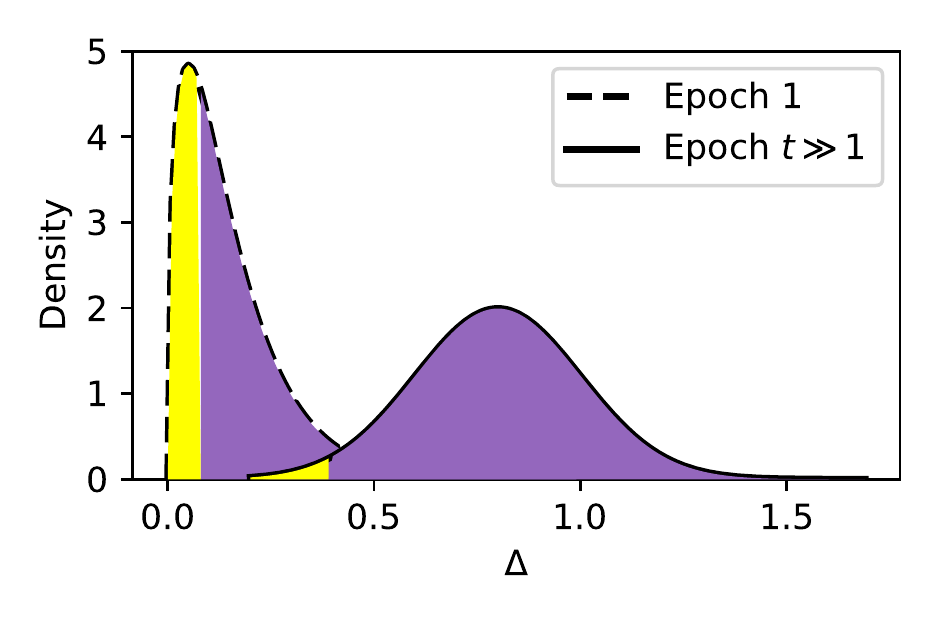}}
		\IfFileExists{figures/automargin/distribution_phi_an-crop.pdf}{}{\immediate\write18{pdfcrop 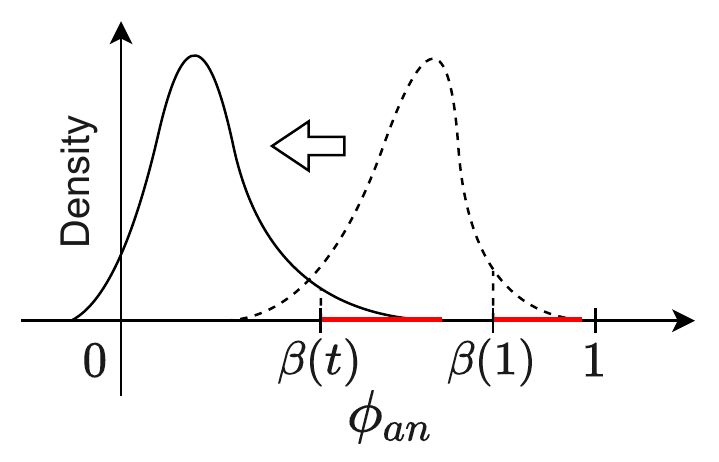}}
		\IfFileExists{figures/mAP-eps-crop.pdf}{}{\immediate\write18{pdfcrop 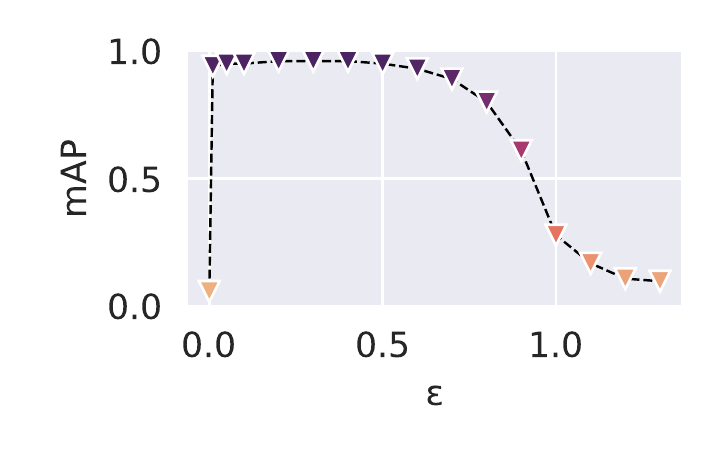}}
		
		\hspace*{\fill}
		\subfloat[\small Effect of $\varepsilon$ in Eq.~\eqref{eq:triplet_loss_norm}\label{fig:eps_mAP_plot}]{ 
\begin{tikzpicture}

\definecolor{burlywood236175128}{RGB}{236,175,128}
\definecolor{darkgray176}{RGB}{176,176,176}
\definecolor{darksalmon234145110}{RGB}{234,145,110}
\definecolor{darksalmon235162120}{RGB}{235,162,120}
\definecolor{darksalmon236165121}{RGB}{236,165,121}
\definecolor{darkslateblue8237100}{RGB}{82,37,100}
\definecolor{darkslateblue9239103}{RGB}{92,39,103}
\definecolor{indigo743498}{RGB}{74,34,98}
\definecolor{indigo763598}{RGB}{76,35,98}
\definecolor{indigo793699}{RGB}{79,36,99}
\definecolor{mediumvioletred16757110}{RGB}{167,57,110}
\definecolor{purple11644110}{RGB}{116,44,110}
\definecolor{salmon22911595}{RGB}{229,115,95}

\begin{axis}[
width=0.37\textwidth,
height=0.27\textwidth,
font=\fontsize{7}{7}\selectfont,
legend style={fill opacity=0.8, draw opacity=1, text opacity=1, draw=none},
tick align=outside,
tick pos=left,
x grid style={darkgray176},
xlabel={$\varepsilon$},
xmajorgrids,
xmin=-0.065, xmax=1.365,
xtick style={color=black},
y grid style={darkgray176},
ylabel={mAP},
ymajorgrids,
ymin=0, ymax=1,
ytick style={color=black},
ytick={0,0.5,1},
yticklabels={0,0.5,1}
]
\addplot [line width=0.28pt, black, dashed, forget plot]
table {%
0 0.06
0.01 0.94
0.05 0.95
0.1 0.95
0.2 0.96
0.3 0.96
0.4 0.96
0.5 0.95
0.6 0.93
0.7 0.89
0.8 0.8
0.9 0.61
1 0.28
1.1 0.17
1.2 0.11
1.3 0.1
};
\addplot [semithick, burlywood236175128, mark=triangle*, mark size=3, mark options={solid,rotate=180,draw=white}, only marks, forget plot]
table {%
0 0.06
};
\addplot [semithick, darksalmon236165121, mark=triangle*, mark size=3, mark options={solid,rotate=180,draw=white}, only marks, forget plot]
table {%
1.3 0.1
};
\addplot [semithick, darksalmon235162120, mark=triangle*, mark size=3, mark options={solid,rotate=180,draw=white}, only marks, forget plot]
table {%
1.2 0.11
};
\addplot [semithick, darksalmon234145110, mark=triangle*, mark size=3, mark options={solid,rotate=180,draw=white}, only marks, forget plot]
table {%
1.1 0.17
};
\addplot [semithick, salmon22911595, mark=triangle*, mark size=3, mark options={solid,rotate=180,draw=white}, only marks, forget plot]
table {%
1 0.28
};
\addplot [semithick, mediumvioletred16757110, mark=triangle*, mark size=3, mark options={solid,rotate=180,draw=white}, only marks, forget plot]
table {%
0.9 0.61
};
\addplot [semithick, purple11644110, mark=triangle*, mark size=3, mark options={solid,rotate=180,draw=white}, only marks, forget plot]
table {%
0.8 0.8
};
\addplot [semithick, darkslateblue9239103, mark=triangle*, mark size=3, mark options={solid,rotate=180,draw=white}, only marks, forget plot]
table {%
0.7 0.89
};
\addplot [semithick, darkslateblue8237100, mark=triangle*, mark size=3, mark options={solid,rotate=180,draw=white}, only marks, forget plot]
table {%
0.6 0.93
};
\addplot [semithick, indigo793699, mark=triangle*, mark size=3, mark options={solid,rotate=180,draw=white}, only marks, forget plot]
table {%
0.01 0.94
};
\addplot [semithick, indigo763598, mark=triangle*, mark size=3, mark options={solid,rotate=180,draw=white}, only marks, forget plot]
table {%
0.05 0.95
0.1 0.95
0.5 0.95
};
\addplot [semithick, indigo743498, mark=triangle*, mark size=3, mark options={solid,rotate=180,draw=white}, only marks, forget plot]
table {%
0.2 0.96
0.3 0.96
0.4 0.96
};
\end{axis}

\end{tikzpicture}}
		\hfill
		\subfloat[\small Distribution of $\Delta$\label{fig:distribution_delta}]{ \input{figures/tikz/distribution_delta.tex}}
		\hfill
		\subfloat[\small Distribution of $\phi_{an}$\label{fig:distribution_phi_an}]{ \input{figures/tikz/distribution_phi_an}}
		
		
		\hspace*{\fill}
		\caption{\small (a) The sensitivity of the Triplet loss~\eqref{eq:triplet_loss_norm} with the change of $\varepsilon$. (b-c) The convergences of distributions of $\Delta=\phi_{ap}-\phi_{an}$ and $\phi_{an}$ under our loss. Yellow and blue areas, specified by Eqs.~\eqref{eq:automargin_delta} and~\eqref{eq:automargin_phi_an}, indicate hard triplets and hard negative pairs, respectively. 
		}
		\label{fig:delta_phi_distribution}
		\vspace{-3mm}
	\end{figure}
	
	Instead of depending on L2 distances between feature vectors as in~\eqref{eq:triplet_loss}, the TL in Eq.~\eqref{eq:triplet_loss_norm} becomes a function of the cosine similarities $\phi_{ap}$ and $\phi_{an}$ (i.e.\ $\cos (\bm{f}_a, \bm{f}_p)$ and $\cos (\bm{f}_a, \bm{f}_n)$, respectively). In~\cref{fig:gradient_field_triplet}, we graphically demonstrate the 2D loss surface of the TL~\eqref{eq:triplet_loss_norm} with $\varepsilon=0.25$, treating $\phi_{ap}$ and $\phi_{an}$ as its arguments. 
	
	\subsection{Adaptive Gradient Triplet Loss}
	
	The TL in Eq.~\eqref{eq:triplet_loss_norm} only aims to ensure that the distance between the feature vectors $\bm{f}_a$ and $\bm{f}_p$ is strictly less than the distance between the anchor and a negative $\bm{f}_n$.
	Such a formulation, however, allows for the existence of an unexpected scenario where both the distances are arbitrarily small. We present a simple intuition of the scenario in~\Cref{fig:ucircle}.
	Although increasing the margin $\varepsilon$ should enlarge the distance of negative pairs, our empirical evidence in~\cref{fig:eps_mAP_plot} shows that using $\varepsilon > 0.5$ results in a significant drop in performance. Therefore, we propose to explicitly set a threshold on the \textit{virtual angle} between $\bm{f}_a$ and $\bm{f}_n$, that is $\angle(\bm{f}_a, \bm{f}_n) \ge \alpha$, where $\alpha \in [0,\pi / 2]$, which is equivalent to $\cos (\bm{f}_a, \bm{f}_n) - \cos(\alpha) \leq 0$.
	To enforce such a constraint, we minimize the following loss
	\begin{equation}
	\mathcal{L}_{\mathrm{an}} = \left [\phi_{an} - \beta \right ]_{+},
	\label{eq:Lan_regularizer}
	\end{equation}
	where $\beta=\cos(\alpha) \in [0, 1]$. Using this additional term, we introduce \emph{an adaptive gradient triplet loss}, named AdaTriplet, that is a combination of $\mathcal{L}^{*}_{\mathrm{Triplet}}$ and $\mathcal{L}_{\mathrm{an}}$:
	\begin{equation}
	\mathcal{L}_{\mathrm{AdaTriplet}} = \left [\phi_{an} - \phi_{ap}+ \varepsilon \right ]_{+} + \lambda\left [\phi_{an} - \beta \right ]_{+},
	\label{eq:our_triplet_loss_norm}
	\end{equation}
	where $\lambda \in \mathbb{R}_{+}$ is a coefficient, $\varepsilon \in [0, 2)$ is a strict margin, and $\beta \in [0, 1]$ is a relaxing margin.
	\begin{proposition}
		Consider $\|\bm{f}_a\|_2 = \|\bm{f}_p\|_2 = \|\bm{f}_n\|_2 = 1$. Compared to the Triplet loss, the gradients of AdaTriplet w.r.t. $\phi_{ap}$ and $\phi_{an}$ adapt the magnitude and the direction depending on the triplet hardness:
		\begin{align}
		\left(\frac{\partial \mathcal{L}_{\mathrm{AdaTriplet}}(\tau)}{\partial \phi_{ap}} , \frac{\partial \mathcal{L}_{\mathrm{AdaTriplet}}(\tau)}{\partial \phi_{an}}\right) = \left\{
		\begin{matrix}
		(-1, 1+\lambda) & \mathrm{if}\ \tau \in \mathcal{T}_{+} \cap  \mathcal{P}_{+}  \\ 
		(0, \lambda) & \mathrm{if}\ \tau \in (\mathcal{T} \textbackslash \mathcal{T}_{+}) \cap  \mathcal{P}_{+}  \\
		(-1, 1) & \mathrm{if}\ \tau \in \mathcal{T}_{+} \cap  (\mathcal{T} \textbackslash \mathcal{P}_{+})  \\ 
		(0, 0) & \mathrm{otherwise}
		\end{matrix},\right.
		\label{eq:adatriplet_gradients}
		\end{align}
		where $\mathcal{T}_{+}=\{(\bm{x}_a, \bm{x}_p, \bm{x}_n) \mid \phi_{an} - \phi_{ap} + \varepsilon > 0\}$ and $\mathcal{P}_{+}=\{(\bm{x}_a, \bm{x}_p, \bm{x}_n) \mid \phi_{an} > \beta \}$.
		\begin{proof} See~\Cref{proof:prop2}.
		\end{proof}
	\end{proposition}
	
	In~\cref{fig:gradient_field_ours}, we illustrate the negative gradient field of AdaTriplet with $\varepsilon=0.25, \beta=0.1$, and $\lambda = 1$. As such, the 2D coordinate is partitioned into $4$ sub-domains, corresponding to Eq.~\eqref{eq:adatriplet_gradients}. The main distinction of the AdaTriplet loss compared to the TL is that our loss has different gradients depending on the difficulty of hard negative samples. In particular, it enables the optimization of easy triplets with $\phi_{an} > \beta$, which addresses the drawback of TL.
	
	\subsection{AutoMargin: Adaptive Hard Negative Mining}
	Hard negative samples are those where feature space mapping  $f_\theta(\cdot)$ fails to capture semantic similarity between samples. Notably, they have recently been shown to benefit the learning process~\cite{xuan2020hard}. The prior work~\cite{xuan2020hard}, considered incorporating an additional term that is minimized when a hard negative example is detected. Otherwise, the normal TL is minimized. We argue that while this direction is promising, optimizing two different losses for different batches may lead to degenerate behaviour during the optimization process. 
	
	In AdaTriplet, instead of defining hard negatives as the ones for which $\phi_{an} > \phi_{ap}$, we have enforced the numerical constraint on the value of $\phi_{an}$ itself. Empirically, one can observe that this constraint becomes easier to satisfy as we train the model for longer.
	
	Let $\Delta=\phi_{ap} - \phi_{an}$, we rewrite~\eqref{eq:our_triplet_loss_norm} as $\mathcal{L}_{\mathrm{AdaTriplet}} = \left [\varepsilon - \Delta\right ]_{+} + \lambda\left [\phi_{an} - \beta \right ]_{+}$. During the convergence of a model under our loss, the distributions of $\Delta$ and $\phi_{an}$ are supposed to transform as illustrated in~\cref{fig:distribution_delta,fig:distribution_phi_an}, respectively. Here, we propose adjusting the margins $\varepsilon$ and $\beta$ according to the summary statistics of the $\Delta$ and $\phi_{an}$ distributions during the training:
	
	\noindent\begin{tabularx}{\textwidth}{@{}XX@{}}
		\begin{equation}
		\varepsilon(t) = \frac{\mu_{\Delta}(t)}{K_{\Delta}}, \label{eq:automargin_delta}
		\end{equation} &
		\begin{equation}
		\beta(t) = 1 + \frac{\mu_{an}(t) - 1}{K_{an}}, \label{eq:automargin_phi_an}
		\end{equation}
	\end{tabularx}
	where $\mu_{\Delta}(t)$ and $\mu_{an}(t)$ are the means of $\{ \Delta \mid (\bm{x}_a, \bm{x}_p, \bm{x}_n) \in \mathcal{T} \}$ and $\{ \phi_{an} \mid (\bm{x}_a, \bm{x}_p, \bm{x}_n) \in \mathcal{T} \}$ respectively, and $K_\Delta, K_{an}\in \mathbb{Z}_{+}$ are hyperparameters.
	
	The difference in $\varepsilon(t)$ and $\beta(t)$ can be observed from their definition: we aim to enforce the triplet constraint with the highest possible margin, and this progressively raises it. Simultaneously, we want to increase the virtual thresholding angle between anchors and negative samples, which leads to the decrease of $\beta(t)$. We provide a graphical illustration of adaptive margins in~\cref{fig:distribution_delta,fig:distribution_phi_an} using yellow and blue colors, respectively. 
	
	\begin{figure}[t!]
		\centering
		\IfFileExists{figures/dam/Delta_changes-crop.pdf}{}{\immediate\write18{pdfcrop 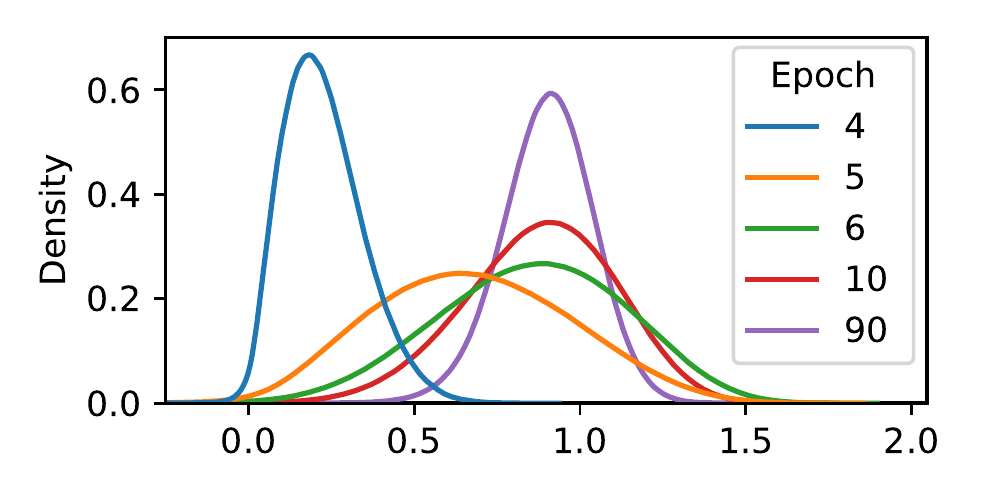}}
		\IfFileExists{figures/automargin/margin_loss_training-crop.pdf}{}{\immediate\write18{pdfcrop 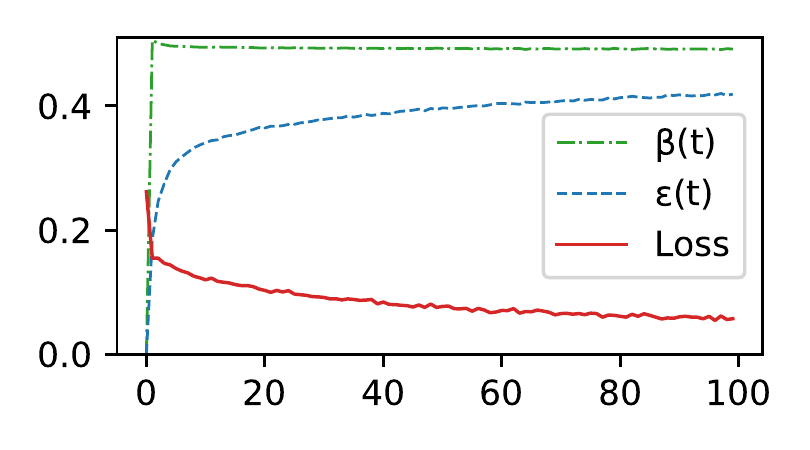}}
		\hspace*{\fill}
		\subfloat[\small Margins and our training loss\label{fig:margins_loss_training_convergence}]{
\begin{tikzpicture}

\definecolor{crimson2143940}{RGB}{214,39,40}
\definecolor{darkgray176}{RGB}{176,176,176}
\definecolor{forestgreen4416044}{RGB}{44,160,44}
\definecolor{lightgray204}{RGB}{204,204,204}
\definecolor{steelblue31119180}{RGB}{31,119,180}

\begin{axis}[
width=0.55\textwidth,
height=0.28\textwidth,
font=\fontsize{6}{6}\selectfont,
xlabel=Epoch,
legend style={fill opacity=0.8, draw opacity=1, text opacity=1, draw=lightgray204},
legend cell align={left},
legend style={
  fill opacity=0.8,
  draw opacity=1,
  text opacity=1,
  at={(.99,0.15)},
  anchor=south east,
  draw=lightgray204
},
tick align=outside,
tick pos=left,
x grid style={darkgray176},
xmin=-4.95, xmax=103.95,
xtick style={color=black},
xtick={0,20,40,60,80,100},
y grid style={darkgray176},
ymin=0, ymax=0.51,
ytick={0,0.25,0.5},
ytick style={color=black}
]
\pgfplotsset{
compat=1.11,
legend image code/.code={
\draw[mark repeat=2,mark phase=2]
plot coordinates {
(0cm,0cm)
(0.1cm,0cm)        
(0.4cm,0cm)         
};%
}
}
\addplot [semithick, forestgreen4416044, dashed]
table {%
0 0
1 0.505974531173706
2 0.500279426574707
4 0.49643886089325
6 0.495498776435852
8 0.494847655296326
10 0.494261741638184
11 0.494221925735474
12 0.494823932647705
13 0.494350671768188
18 0.493755102157593
19 0.493240833282471
28 0.493083953857422
29 0.492714762687683
31 0.493098974227905
32 0.492444753646851
33 0.493092060089111
36 0.492607474327087
37 0.49213719367981
38 0.492872476577759
40 0.492239952087402
41 0.492895722389221
43 0.492063999176025
44 0.492541432380676
48 0.492158770561218
49 0.492956042289734
51 0.492055892944336
54 0.492515802383423
55 0.491688251495361
56 0.492414712905884
57 0.492526054382324
58 0.491451859474182
59 0.491884589195251
60 0.491508364677429
61 0.492188930511475
63 0.492295026779175
64 0.490817904472351
65 0.492066860198975
66 0.491562366485596
68 0.49221658706665
69 0.491511106491089
72 0.491652488708496
73 0.491312980651855
74 0.492157220840454
75 0.491455793380737
77 0.491814136505127
78 0.491257905960083
79 0.492523074150085
80 0.491506218910217
81 0.491500854492188
82 0.490850687026978
84 0.4920654296875
85 0.492113471031189
86 0.491637706756592
87 0.491732954978943
88 0.490982294082642
89 0.491418600082397
90 0.491088151931763
91 0.491495609283447
96 0.49175500869751
97 0.490562796592712
98 0.491876244544983
99 0.491455316543579
};
\addlegendentry{$\beta(t)$}
\addplot [semithick, steelblue31119180, dashed]
table {%
0 0
1 0.185909032821655
2 0.246919751167297
3 0.274498701095581
4 0.297030210494995
5 0.309945821762085
6 0.318137168884277
8 0.332376718521118
9 0.337005615234375
11 0.344170331954956
12 0.345267295837402
13 0.350227355957031
14 0.352231860160828
15 0.353142023086548
17 0.358927249908447
18 0.361723661422729
19 0.365192294120789
20 0.364330172538757
21 0.367255568504333
23 0.367987155914307
24 0.370218515396118
25 0.370264053344727
26 0.372678160667419
27 0.373644232749939
28 0.375155925750732
29 0.377405166625977
30 0.378213763237
32 0.381069421768188
33 0.380931735038757
34 0.383654952049255
35 0.381843090057373
36 0.383589863777161
37 0.386256337165833
38 0.384576320648193
39 0.385821104049683
40 0.387948751449585
41 0.387140870094299
43 0.391441702842712
44 0.391679763793945
46 0.394617557525635
47 0.391996622085571
48 0.395812749862671
49 0.393961310386658
50 0.396760940551758
51 0.396208763122559
52 0.396313667297363
53 0.39765453338623
54 0.398410320281982
55 0.39961314201355
56 0.40032970905304
57 0.399839401245117
59 0.403542995452881
60 0.404027223587036
63 0.402610421180725
64 0.406082034111023
65 0.405319094657898
66 0.405825734138489
67 0.405353903770447
68 0.406248688697815
69 0.406561613082886
70 0.407898187637329
71 0.408623695373535
72 0.407814621925354
73 0.410300612449646
74 0.409008264541626
75 0.410358428955078
76 0.409555554389954
77 0.409502983093262
78 0.412544965744019
79 0.411185383796692
80 0.413219213485718
81 0.413855314254761
82 0.415390133857727
84 0.413304209709167
85 0.412604928016663
86 0.413918852806091
87 0.413817644119263
88 0.417747974395752
89 0.416609764099121
90 0.417775392532349
92 0.415964245796204
93 0.416468739509583
94 0.416436433792114
95 0.418224573135376
96 0.417527556419373
97 0.419844627380371
98 0.416924953460693
99 0.418853521347046
};
\addlegendentry{$\varepsilon(t)$}
\addplot [semithick, crimson2143940]
table {%
0 0.261754631996155
1 0.155001878738403
2 0.154868960380554
3 0.146625280380249
4 0.14411735534668
5 0.138247847557068
6 0.134132266044617
7 0.131190061569214
8 0.125942707061768
9 0.123364567756653
10 0.120081663131714
11 0.122862458229065
12 0.117634654045105
14 0.115012049674988
15 0.112388134002686
16 0.110854506492615
17 0.110910177230835
18 0.109234929084778
19 0.10525643825531
20 0.102969527244568
21 0.0999305248260498
22 0.103084564208984
23 0.100514531135559
24 0.102672696113586
25 0.0969825983047485
27 0.0951337814331055
28 0.0930558443069458
29 0.0926403999328613
30 0.0917139053344727
31 0.0894286632537842
32 0.0896024703979492
33 0.0877314805984497
34 0.0894142389297485
35 0.0884585380554199
36 0.0870529413223267
37 0.0874315500259399
38 0.0884853601455688
39 0.0813469886779785
40 0.0843244791030884
41 0.0804572105407715
42 0.0802333354949951
43 0.0790615081787109
44 0.0785075426101685
45 0.0762647390365601
46 0.079625129699707
47 0.0756804943084717
48 0.0810712575912476
49 0.0756381750106812
50 0.0773040056228638
51 0.0778377056121826
52 0.0735466480255127
53 0.0734298229217529
54 0.0741091966629028
55 0.0695598125457764
56 0.074036717414856
57 0.0716592073440552
58 0.0671429634094238
59 0.0681962966918945
60 0.0707792043685913
61 0.0706225633621216
62 0.0737978219985962
63 0.0664441585540771
64 0.0690511465072632
65 0.0685975551605225
66 0.0713851451873779
68 0.0678305625915527
69 0.0636175870895386
70 0.0658166408538818
71 0.0661520957946777
72 0.0647083520889282
73 0.0658760070800781
74 0.0639606714248657
75 0.0664503574371338
76 0.0658013820648193
77 0.0598636865615845
78 0.0633955001831055
79 0.0629440546035767
81 0.0600374937057495
82 0.0646175146102905
83 0.0613949298858643
84 0.0654979944229126
86 0.0599832534790039
87 0.0570821762084961
88 0.0587639808654785
89 0.0581868886947632
90 0.0605493783950806
91 0.0614653825759888
92 0.0600788593292236
93 0.0599048137664795
94 0.05733323097229
95 0.061303973197937
96 0.054714560508728
97 0.0619616508483887
98 0.0561422109603882
99 0.0575699806213379
};
\addlegendentry{Loss}
\end{axis}

\end{tikzpicture}}
		\hfill
		\subfloat[\small Evolution of distribution of $\Delta$\label{fig:phi_deltal_training_convergence}]{\input{figures/tikz/Delta_changes}}
		\hspace*{\fill}
		\caption{\small Effects of AdaTriplet and AutoMargin. Colors in (b) represent epochs.}
		\vspace{-3mm}
		\label{fig:delta_phi_distribution_convergence}
	\end{figure}
	
	\section{Experiments}
	
	\subsection{Datasets}
	\noindent\textbf{Knee X-ray dataset.}
	The Osteoarthritis Initiative (OAI) cohort, publicly available at \url{https://nda.nih.gov/oai/}, comprises $4,796$ participants from $45$ to $79$ years old. The original interest of the cohort was to study knee osteoarthritis, which is characterized by the appearance of osteophytes, joint space narrowing, as well as textural changes of the femur and tibia. 
	We used X-ray imaging data collected at baseline, $12$, $24$, $36$, $48$, $72$, and $96$-month follow-up visits.
	The detailed data description is presented in~\Cref{tbl:data_descriptions}. 
	We utilized KNEEL~\cite{tiulpin2019kneel} to localize and crop a pair of knees joints from each bilateral radiograph. Our further post-processing used augmentations that eventually produces input images with a shape of $256\times256$ (see \Cref{tbl:oai_train_augmentation} for details). 
	
	\noindent\textbf{Chest X-ray dataset.} 
	ChestXrays-14 (CXR)~\cite{wang2017chestx} consists of $112,120$ frontal-view chest X-ray images collected from $30,805$ participants from $0$ to $95$ years old. The radiographic data were acquired at a baseline and across time up to $156$ months. The training and test data are further described in~\Cref{tbl:data_descriptions}. To be in line with the OAI dataset, we grouped testing data by year, and used the same set of augmentations, yielding $256\times256$ images. 
	
	\subsection{Experimental Setup}
	We conducted our experiments on V100 Nvidia GPUs. We implemented our method and all baselines in PyTorch~\cite{paszke2019pytorch} and the Metric Learning library~\cite{musgrave2020pytorch}. Following~\cite{musgrave2020metric}, the same data settings, optimizer hyperparameters, augmentations, and feature extraction module were used for all the methods. We utilized the Adam optimizer~\cite{kingma2014adam} with a learning rate of $0.0001$ and a weight decay of $0.0001$. We used the ResNet-18 network~\cite{he2016deep} with pretrained weights to extract embeddings with $D$ of $128$ from input images. We trained each method in $100$ epochs with a batch size of $128$. For data sampling in each batch, we randomly selected $4$ medical images from each subject. We thoroughly describe lists of hyperparameters for all the methods in~\Cref{tbl:hyperparameter_range}.
	
	To evaluate forensic matching performance, we used mean average precision (mAP)~\cite{schutze2008introduction}, mAP@R~\cite{musgrave2020metric}, and cumulative matching characteristics (CMC) accuracy~\cite{decann2013relating}. All experiments were run $5$ times with different random seeds. All test set metrics represent the average and standard error over runs.

	\begin{table}[t!]
		\caption{\small Ablation studies ($5$-fold CV; OAI dataset). CMC means CMC top-1. $^*$ indicates the results when the query and the database are $6$ years apart. $N_s$ is the number of scanned hyperparameter values.}
		\centering
		\subfloat[Impact of $\lambda\mathcal{L}_{\mathrm{an}}$\label{tbl:lambda_scanning}]{
			\scalebox{0.8}{
				\begin{tabular}{ccccc}
					\toprule
					\textBF{$\lambda$}         &\textBF{mAP$^*$}            & \textBF{CMC$^*$}  & \textBF{mAP}            & \textBF{CMC}    \\ \midrule
					0 &      95.6 &      93.4 &      96.6 &      93.6 \\
					
					0.5 &      96.1 &      94.4 &      96.9 &      94.6 \\
					
					\textBF{1} &      \textBF{96.3} &      \textBF{94.6} &      \textBF{97.0} &      \textBF{94.7} \\
					
					2 &      94.5 &      92.1 &      95.6 &      92.3 \\
					\bottomrule
				\end{tabular}
		}}
		\quad
		\subfloat[Triplet loss\label{tbl:ablation_automargin_triplet}]{
			\scalebox{0.8}{
				\begin{tabular}{lccc}
					\toprule
					\textBF{Method} & $N_s$ & \textBF{mAP} & \textBF{CMC}       \\
					\midrule
					Q1  & 1 & 27.3 & 14.9 \\
					Q2 & 1 & 87.7 & 76.9 \\
					WAT~\cite{zhao2019weakly} & 4 &96.5 & 93.5 \\
					Grid search & 4 & \textBF{96.6} & 93.6\\
					AutoMargin & 2 & \textBF{96.6} & \textBF{93.7} \\
					\bottomrule
		\end{tabular} }}
		\quad
		\subfloat[AdaTriplet loss\label{tbl:ablation_automargin}]{
			\scalebox{0.8}{
				\begin{tabular}{lccc}
					\toprule
					\textBF{Method} & $N_s$ & \textBF{mAP}  & \textBF{CMC}      \\
					\midrule
					Q1 & 1 & 94.3 & 89.4 \\
					Q2 & 1 &88.9 & 79.5 \\
					Grid search & 16 & 97.0 & \textBF{94.8} \\
					AutoMargin & 4 & \textBF{97.1} & 94.7 \\
					\bottomrule
		\end{tabular} }}
		\vspace{-6mm}
	\end{table}
	
	\subsection{Results}
	
	\noindent\textbf{Impact of $\mathcal{L}_{an}$.}
	We performed an experiment in which we varied the coefficient $\lambda$ in the AdaTriplet loss~\eqref{eq:our_triplet_loss_norm}. The results on the OAI test set in~\cref{tbl:lambda_scanning} show that $\lambda=1$  yielded the best performances according to both the mAP and CMC metrics. Notably, we observed that the differences are more apparent when querying images at least $6$ years apart from images in the database. We thus set $\lambda=1$ for our method in all other experiments.
	
	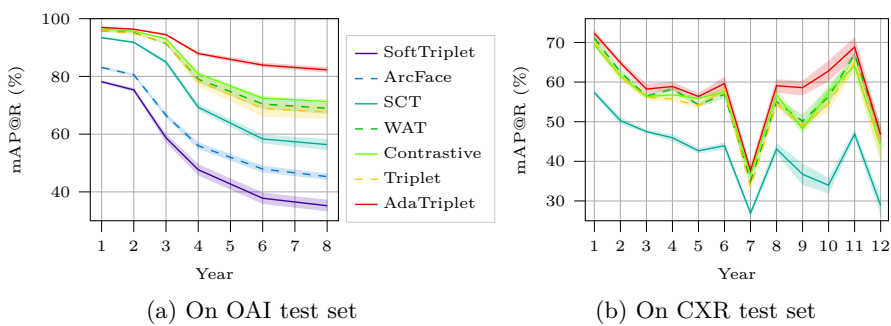
\begin{figure}[t!]
		\centering
		\hspace*{\fill}
		\subfloat[\small On OAI test set\label{fig:oai_performances}]{
\begin{tikzpicture}

\definecolor{darkcyan0170151}{RGB}{0,170,151}
\definecolor{darkgray176}{RGB}{176,176,176}
\definecolor{dodgerblue0119221}{RGB}{0,119,221}
\definecolor{gold2552010}{RGB}{255,201,0}
\definecolor{green01880}{RGB}{0,188,0}
\definecolor{indigo660161}{RGB}{66,0,161}
\definecolor{lawngreen1022550}{RGB}{102,255,0}
\definecolor{lightgray204}{RGB}{204,204,204}
\definecolor{red23500}{RGB}{235,0,0}

\begin{axis}[
width=0.4\textwidth,
height=0.35\textwidth,
font=\fontsize{6}{6}\selectfont,
legend cell align={left},
legend style={
  fill opacity=0.8,
  draw opacity=1,
  text opacity=1,
  at={(1.03,-0.01)},
  anchor=south west,
  draw=lightgray204
},
tick align=outside,
tick pos=left,
x grid style={darkgray176},
xlabel={Year},
xmajorgrids,
xmin=0.65, xmax=8.35,
xtick style={color=black},
xtick={1,2,3,4,5,6,7,8},
y grid style={darkgray176},
ylabel={mAP@R (\%)},
ymajorgrids,
ymin=30, ymax=100,
ytick style={color=black}
]

\pgfplotsset{
compat=1.11,
legend image code/.code={
\draw[mark repeat=2,mark phase=2]
plot coordinates {
(0cm,0cm)
(0.1cm,0cm)        
(0.3cm,0cm)         
};%
}
}

\addlegendentry{SoftTriplet}
\addlegendentry{ArcFace}
\addlegendentry{SCT}
\addlegendentry{WAT}
\addlegendentry{Contrastive}
\addlegendentry{Triplet}
\addlegendentry{AdaTriplet}

\path [draw=indigo660161, fill=indigo660161, opacity=0.2]
(axis cs:1,78.4595352564102)
--(axis cs:1,77.8545673076923)
--(axis cs:2,74.8558484349258)
--(axis cs:3,57.3936631944444)
--(axis cs:4,45.9301333333333)
--(axis cs:6,36.0183189655172)
--(axis cs:8,33.4367612293144)
--(axis cs:8,37.0951536643026)
--(axis cs:8,37.0951536643026)
--(axis cs:6,39.5689655172413)
--(axis cs:4,49.42)
--(axis cs:3,59.7894965277777)
--(axis cs:2,75.743410214168)
--(axis cs:1,78.4595352564102)
--cycle;

\path [draw=dodgerblue0119221, fill=dodgerblue0119221, opacity=0.2]
(axis cs:1,83.3373397435897)
--(axis cs:1,82.8505608974358)
--(axis cs:2,80.2780065897858)
--(axis cs:3,65.9809027777777)
--(axis cs:4,55.0377777777778)
--(axis cs:6,46.931573275862)
--(axis cs:8,44.3705673758865)
--(axis cs:8,46.1998817966903)
--(axis cs:8,46.1998817966903)
--(axis cs:6,48.9197198275862)
--(axis cs:4,56.9666666666667)
--(axis cs:3,67.1050347222222)
--(axis cs:2,80.7987644151565)
--(axis cs:1,83.3373397435897)
--cycle;

\path [draw=darkcyan0170151, fill=darkcyan0170151, opacity=0.2]
(axis cs:1,93.6598557692307)
--(axis cs:1,93.1770833333333)
--(axis cs:2,91.6410214168039)
--(axis cs:3,84.2469618055555)
--(axis cs:4,68.4422222222222)
--(axis cs:6,56.9612068965517)
--(axis cs:8,54.6306146572103)
--(axis cs:8,58.1294326241134)
--(axis cs:8,58.1294326241134)
--(axis cs:6,59.665948275862)
--(axis cs:4,70.2866666666666)
--(axis cs:3,85.6618923611111)
--(axis cs:2,91.9563426688632)
--(axis cs:1,93.6598557692307)
--cycle;

\path [draw=green01880, fill=green01880, opacity=0.2]
(axis cs:1,95.7852564102563)
--(axis cs:1,95.6911057692307)
--(axis cs:2,95.1132619439868)
--(axis cs:3,91.2717013888888)
--(axis cs:4,77.9708444444444)
--(axis cs:6,68.9682112068965)
--(axis cs:8,67.2074468085106)
--(axis cs:8,70.5851063829787)
--(axis cs:8,70.5851063829787)
--(axis cs:6,72.020474137931)
--(axis cs:4,80.1555555555555)
--(axis cs:3,91.5625)
--(axis cs:2,95.2553542009884)
--(axis cs:1,95.7852564102563)
--cycle;

\path [draw=lawngreen1022550, fill=lawngreen1022550, opacity=0.2]
(axis cs:1,96.5404647435897)
--(axis cs:1,96.3040865384615)
--(axis cs:2,95.6257001647446)
--(axis cs:3,92.8927951388888)
--(axis cs:4,80.3182222222222)
--(axis cs:6,71.9127155172414)
--(axis cs:8,70.6205673758865)
--(axis cs:8,71.9444444444444)
--(axis cs:8,71.9444444444444)
--(axis cs:6,73.0980603448276)
--(axis cs:4,81.4777777777778)
--(axis cs:3,93.0971354166666)
--(axis cs:2,95.7289950576606)
--(axis cs:1,96.5404647435897)
--cycle;

\path [draw=gold2552010, fill=gold2552010, opacity=0.2]
(axis cs:1,95.9334935897435)
--(axis cs:1,95.6670673076922)
--(axis cs:2,94.7549423393739)
--(axis cs:3,91.4214409722222)
--(axis cs:4,76.8422222222222)
--(axis cs:6,66.2526939655172)
--(axis cs:8,65.4078014184397)
--(axis cs:8,69.9940898345153)
--(axis cs:8,69.9940898345153)
--(axis cs:6,70.9859913793103)
--(axis cs:4,80.0022222222222)
--(axis cs:3,92.0724826388888)
--(axis cs:2,95.0638385502471)
--(axis cs:1,95.9334935897435)
--cycle;

\path [draw=red23500, fill=red23500, opacity=0.2]
(axis cs:1,97.0572916666666)
--(axis cs:1,96.8569711538461)
--(axis cs:2,96.24176276771)
--(axis cs:3,94.314236111111)
--(axis cs:4,87.4444444444444)
--(axis cs:6,83.2812499999999)
--(axis cs:8,81.5159574468085)
--(axis cs:8,83.0378250591016)
--(axis cs:8,83.0378250591016)
--(axis cs:6,84.598599137931)
--(axis cs:4,88.3644444444444)
--(axis cs:3,94.6050347222222)
--(axis cs:2,96.4394563426688)
--(axis cs:1,97.0572916666666)
--cycle;

\addplot [line width=0.52pt, indigo660161]
table {%
1 78.1610565185547
2 75.333610534668
3 58.7152786254883
4 47.6666679382324
6 37.793643951416
8 35.2038993835449
};
\addplot [line width=0.52pt, dodgerblue0119221, dashed]
table {%
1 83.1029663085938
2 80.5663070678711
3 66.5277786254883
4 55.9799995422363
6 47.9256477355957
8 45.2807312011719
};
\addplot [line width=0.52pt, darkcyan0170151]
table {%
1 93.4174652099609
2 91.7936553955078
3 84.9544296264648
4 69.3088912963867
6 58.340518951416
8 56.3800239562988
};
\addplot [line width=0.52pt, green01880, dashed]
table {%
1 95.7371826171875
2 95.1812210083008
3 91.4344635009766
4 79.0400009155273
6 70.431037902832
8 68.9598083496094
};
\addplot [line width=0.52pt, lawngreen1022550]
table {%
1 96.4202728271484
2 95.679573059082
3 92.9947891235352
4 80.8644409179688
6 72.4676742553711
8 71.2825088500977
};
\addplot [line width=0.52pt, gold2552010, dashed]
table {%
1 95.8032836914062
2 94.9093933105469
3 91.7469635009766
4 78.3977813720703
6 68.8469848632812
8 67.7009429931641
};
\addplot [line width=0.52pt, red23500]
table {%
1 96.9591369628906
2 96.3467864990234
3 94.4618072509766
4 87.9044418334961
6 83.9170227050781
8 82.2754135131836
};
\end{axis}

\end{tikzpicture}}
		\hfill
		\subfloat[\small On CXR test set\label{fig:cxr_performances}]{
\begin{tikzpicture}

\definecolor{darkcyan0170151}{RGB}{0,170,151}
\definecolor{darkgray176}{RGB}{176,176,176}
\definecolor{dodgerblue0119221}{RGB}{0,119,221}
\definecolor{gold2552010}{RGB}{255,201,0}
\definecolor{green01880}{RGB}{0,188,0}
\definecolor{indigo660161}{RGB}{66,0,161}
\definecolor{lawngreen1022550}{RGB}{102,255,0}
\definecolor{lightgray204}{RGB}{204,204,204}
\definecolor{red23500}{RGB}{235,0,0}

\begin{axis}[
width=0.46\textwidth,
height=0.35\textwidth,
font=\fontsize{6}{6}\selectfont,
legend cell align={left},
legend style={
  fill opacity=0.8,
  draw opacity=1,
  text opacity=1,
  at={(1.07,-0.01)},
  anchor=south west,
  draw=lightgray204
},
tick align=outside,
tick pos=left,
x grid style={darkgray176},
xlabel={Year},
xmajorgrids,
xmin=0.65, xmax=12.3,
xtick style={color=black},
xtick={1,2,3,4,5,6,7,8,9,10,11,12},
ytick={30, 40, 50, 60, 70},
y grid style={darkgray176},
ylabel={mAP@R (\%)},
ymajorgrids,
ymin=25, ymax=76,
ytick style={color=black}
]
\path [draw=indigo660161, fill=indigo660161, opacity=0.2]
(axis cs:1,15.0017022630638)
--(axis cs:1,14.7744323587303)
--(axis cs:2,10.8084455265959)
--(axis cs:3,8.45836518265912)
--(axis cs:4,7.57373053486381)
--(axis cs:5,7.93204554235636)
--(axis cs:6,7.84687493966217)
--(axis cs:7,4.28409801908371)
--(axis cs:8,5.09831197366037)
--(axis cs:9,7.86555973266499)
--(axis cs:10,3.25301528018614)
--(axis cs:11,3.34965667422414)
--(axis cs:12,0.89050595238095)
--(axis cs:12,1.78170634920635)
--(axis cs:12,1.78170634920635)
--(axis cs:11,5.01655138586903)
--(axis cs:10,4.00625850340136)
--(axis cs:9,10.4021094402673)
--(axis cs:8,6.51836559196066)
--(axis cs:7,4.84611333654513)
--(axis cs:6,8.65542824771071)
--(axis cs:5,8.39104066170346)
--(axis cs:4,8.29657069019182)
--(axis cs:3,9.11285020320858)
--(axis cs:2,11.2777661390433)
--(axis cs:1,15.0017022630638)
--cycle;

\path [draw=dodgerblue0119221, fill=dodgerblue0119221, opacity=0.2]
(axis cs:1,16.8246906399305)
--(axis cs:1,16.5516721356655)
--(axis cs:2,11.9577196172014)
--(axis cs:3,10.066477069702)
--(axis cs:4,8.51077447898707)
--(axis cs:5,8.42435927210652)
--(axis cs:6,9.09075274397522)
--(axis cs:7,5.10127252814112)
--(axis cs:8,6.72237377976475)
--(axis cs:9,8.09155086507058)
--(axis cs:10,4.36014109347442)
--(axis cs:11,4.69862211836606)
--(axis cs:12,1.14399801587301)
--(axis cs:12,1.74583333333333)
--(axis cs:12,1.74583333333333)
--(axis cs:11,5.65329847827425)
--(axis cs:10,5.79158507847583)
--(axis cs:9,10.8339433045025)
--(axis cs:8,7.3676553762097)
--(axis cs:7,5.94473403875996)
--(axis cs:6,9.88967476187204)
--(axis cs:5,9.55383142908039)
--(axis cs:4,9.43414943446682)
--(axis cs:3,10.6759458692893)
--(axis cs:2,12.4878022726083)
--(axis cs:1,16.8246906399305)
--cycle;

\path [draw=darkcyan0170151, fill=darkcyan0170151, opacity=0.2]
(axis cs:1,57.7990319778357)
--(axis cs:1,56.8790185713807)
--(axis cs:2,49.4638087419359)
--(axis cs:3,47.1911545875274)
--(axis cs:4,45.1601386512485)
--(axis cs:5,42.0960685060929)
--(axis cs:6,43.1042303004268)
--(axis cs:7,26.3985873020771)
--(axis cs:8,42.0232922553972)
--(axis cs:9,34.1734591675381)
--(axis cs:10,31.9732266551305)
--(axis cs:11,46.1142727079405)
--(axis cs:12,26.9298492248492)
--(axis cs:12,30.8115958578458)
--(axis cs:12,30.8115958578458)
--(axis cs:11,47.68190805478)
--(axis cs:10,35.7164588730006)
--(axis cs:9,39.2727563511398)
--(axis cs:8,44.4143745455438)
--(axis cs:7,27.3623589547614)
--(axis cs:6,44.6169205955396)
--(axis cs:5,43.1994451641141)
--(axis cs:4,46.7172392706512)
--(axis cs:3,47.7683873170488)
--(axis cs:2,51.0173023765454)
--(axis cs:1,57.7990319778357)
--cycle;

\path [draw=green01880, fill=green01880, opacity=0.2]
(axis cs:1,71.7390833390653)
--(axis cs:1,70.4388617716293)
--(axis cs:2,61.6116114296955)
--(axis cs:3,56.1401232679878)
--(axis cs:4,57.8043010891394)
--(axis cs:5,53.8619322301583)
--(axis cs:6,55.9531258393276)
--(axis cs:7,34.4445287705283)
--(axis cs:8,54.2764853194887)
--(axis cs:9,48.3472662223602)
--(axis cs:10,54.4603492963455)
--(axis cs:11,66.5884785405546)
--(axis cs:12,45.1129785409035)
--(axis cs:12,49.0182495051245)
--(axis cs:12,49.0182495051245)
--(axis cs:11,67.4638674602342)
--(axis cs:10,58.3078442269777)
--(axis cs:9,51.6608381014696)
--(axis cs:8,55.675771658318)
--(axis cs:7,35.4309593398466)
--(axis cs:6,57.8711727917125)
--(axis cs:5,54.7097562222058)
--(axis cs:4,58.6883604857887)
--(axis cs:3,56.8553745100569)
--(axis cs:2,63.381789886603)
--(axis cs:1,71.7390833390653)
--cycle;

\path [draw=lawngreen1022550, fill=lawngreen1022550, opacity=0.2]
(axis cs:1,70.0313123208741)
--(axis cs:1,68.9143209403614)
--(axis cs:2,60.6907983934476)
--(axis cs:3,55.9946734397493)
--(axis cs:4,56.1944248651001)
--(axis cs:5,55.4641429045834)
--(axis cs:6,56.9297671835161)
--(axis cs:7,36.0253006002311)
--(axis cs:8,55.3196726430685)
--(axis cs:9,46.9477847740535)
--(axis cs:10,55.5839021201406)
--(axis cs:11,62.5256345624313)
--(axis cs:12,40.5100658831909)
--(axis cs:12,47.4801623376623)
--(axis cs:12,47.4801623376623)
--(axis cs:11,65.6790864001343)
--(axis cs:10,59.5138433537473)
--(axis cs:9,49.308308102716)
--(axis cs:8,58.3032489478842)
--(axis cs:7,36.8849015405415)
--(axis cs:6,58.1625228159933)
--(axis cs:5,56.3331080770806)
--(axis cs:4,57.3965687253786)
--(axis cs:3,56.661601392612)
--(axis cs:2,62.3336416148376)
--(axis cs:1,70.0313123208741)
--cycle;

\path [draw=gold2552010, fill=gold2552010, opacity=0.2]
(axis cs:1,70.9893703988852)
--(axis cs:1,69.4065234681132)
--(axis cs:2,60.7891404450071)
--(axis cs:3,55.5059922260121)
--(axis cs:4,55.1915750948747)
--(axis cs:5,53.4089042776712)
--(axis cs:6,57.4593493973553)
--(axis cs:7,32.9682820971575)
--(axis cs:8,53.5745428821246)
--(axis cs:9,48.6322343244335)
--(axis cs:10,52.6739795009761)
--(axis cs:11,63.5331427882639)
--(axis cs:12,40.6002934565434)
--(axis cs:12,45.8463628260628)
--(axis cs:12,45.8463628260628)
--(axis cs:11,66.2127333911763)
--(axis cs:10,56.2234609322324)
--(axis cs:9,49.2437786993707)
--(axis cs:8,55.5447305334563)
--(axis cs:7,33.8190529580111)
--(axis cs:6,58.8327799541475)
--(axis cs:5,54.8572034694312)
--(axis cs:4,56.2316457510438)
--(axis cs:3,56.8059200261602)
--(axis cs:2,62.1378783408725)
--(axis cs:1,70.9893703988852)
--cycle;

\path [draw=red23500, fill=red23500, opacity=0.2]
(axis cs:1,72.9485393591911)
--(axis cs:1,71.7122549642612)
--(axis cs:2,64.2906285228278)
--(axis cs:3,57.2073665935814)
--(axis cs:4,58.093288453934)
--(axis cs:5,55.6953409341767)
--(axis cs:6,58.1126013953469)
--(axis cs:7,36.7992033750798)
--(axis cs:8,57.5038673661554)
--(axis cs:9,56.7080528963517)
--(axis cs:10,60.3420591303824)
--(axis cs:11,66.5544119875657)
--(axis cs:12,45.0900267787767)
--(axis cs:12,48.337702020202)
--(axis cs:12,48.337702020202)
--(axis cs:11,71.122011203677)
--(axis cs:10,65.273975849623)
--(axis cs:9,60.1403471771892)
--(axis cs:8,60.5499319727891)
--(axis cs:7,38.5813206240257)
--(axis cs:6,61.1775698662902)
--(axis cs:5,57.0150614857023)
--(axis cs:4,59.9171629765374)
--(axis cs:3,59.3685466079864)
--(axis cs:2,65.4488512874459)
--(axis cs:1,72.9485393591911)
--cycle;

\addplot [line width=0.52pt, indigo660161]
table {%
1 14.8930292129517
2 11.0445365905762
3 8.80223751068115
4 7.9093165397644
5 8.17124557495117
6 8.24071598052979
7 4.57046604156494
8 5.83807134628296
9 9.14765071868896
10 3.64036250114441
11 4.16134166717529
12 1.33668649196625
};
\addplot [line width=0.52pt, dodgerblue0119221, dashed]
table {%
1 16.6849784851074
2 12.2227611541748
3 10.3712110519409
4 8.97091007232666
5 8.98529624938965
6 9.48702144622803
7 5.52071571350098
8 7.03656578063965
9 9.46274662017822
10 5.0883412361145
11 5.17596006393433
12 1.44158399105072
};

\addplot [line width=0.52pt, darkcyan0170151]
table {%
1 57.3348121643066
2 50.2672882080078
3 47.4975357055664
4 45.9504241943359
5 42.6477584838867
6 43.9303894042969
7 26.9105319976807
8 43.1129951477051
9 36.7231063842773
10 33.9640617370605
11 46.9200286865234
12 28.8707218170166
};
\addplot [line width=0.52pt, green01880, dashed]
table {%
1 71.0998992919922
2 62.4967002868652
3 56.4889373779297
4 58.1928939819336
5 54.2858428955078
6 56.9254112243652
7 34.9595794677734
8 54.9761276245117
9 50.1425437927246
10 56.317325592041
11 67.0261764526367
12 47.0144462585449
};
\addplot [line width=0.52pt, lawngreen1022550]
table {%
1 69.4282684326172
2 61.5122184753418
3 56.3424263000488
4 56.7854423522949
5 55.8921737670898
6 57.5118522644043
7 36.4711837768555
8 56.7976875305176
9 48.1280479431152
10 57.5488739013672
11 64.0980758666992
12 43.8911895751953
};
\addplot [line width=0.52pt, gold2552010, dashed]
table {%
1 70.2403259277344
2 61.4536285400391
3 56.1590385437012
4 55.7116088867188
5 54.0841941833496
6 58.1554679870605
7 33.4272880554199
8 54.5596351623535
9 48.9301528930664
10 54.4487190246582
11 64.8729400634766
12 43.3889350891113
};
\addplot [line width=0.52pt, red23500]
table {%
1 72.3303985595703
2 64.869743347168
3 58.2281723022461
4 58.8869972229004
5 56.3686637878418
6 59.6396522521973
7 37.6740951538086
8 59.0832786560059
9 58.585750579834
10 62.8080177307129
11 68.8382110595703
12 46.6864624023438
};
\end{axis}

\end{tikzpicture}}
		\hspace*{\fill}
		\caption{\small Performance comparisons on the test sets of OAI and CXR (mean and standard error over $5$ random seeds). Detailed quantitative results are in~\Cref{tbl:oai_performances,tbl:cxr_performances}.}
		\label{fig:performances}
		\vspace{-3mm}
	\end{figure}

	\noindent\textbf{Impact of AutoMargin.}
	AutoMargin is applicable for both TL and AdaTriplet, and we investigated its impact in  \cref{tbl:ablation_automargin_triplet,tbl:ablation_automargin}. For baselines, we used the Q1 and Q2 quartiles of distributions of $\Delta$ and $\phi_{an}$ to define the margins $\varepsilon$ and $\beta$, respectively. In addition, we performed exhaustive grid searches for the two losses' margins. Besides the na\"ive baselines, we compared our method to the weakly adaptive triplet loss (WAT)~\cite{zhao2019weakly}, which also allows for dynamic margin adjustment in the TL. 
	Based on~\Cref{tbl:grid_search}, we set the constants $(K_\Delta,K_{an})$ of AutoMargin to $(2, 2)$ and $(2, 4)$ for OAI and CXR, respectively.  
	
	AutoMargin helped both the losses to outperform the quartile-based approaches. Compared to the grid search, our method was at least $2$-fold more efficient, and performed in par with the baseline. In the TL, AutoMargin was $2$ time more efficient and achieved better results compared to WAT. Furthermore, on the independent test sets, the combination of AdaTriplet and AutoMargin gained substantially higher performances than WAT (\cref{fig:performances}).

	\noindent\textbf{Effects of our methods in training.}
	We demonstrate the behaviour of AdaTriplet and AutoMargin during training of one of the runs of the OAI experiments in~\cref{fig:delta_phi_distribution_convergence}. Specifically, under our adaptive hard negative mining, the margin $\beta$ drastically increased from $0$ to $0.5$ in a few epochs.
	While $\beta$ was stable after the drastic increase in value, the margin $\varepsilon$ gradually grew from $0$ and converged around $0.4$. As a result, our loss improved rapidly at the beginning, and continuously converged afterwards (see~\cref{fig:margins_loss_training_convergence}). During the process, the mean of $\Delta$ shifted away from $0$ to $1$ while its variance increased at first, and then gradually decreased (\cref{fig:phi_deltal_training_convergence}).

	\noindent\textbf{Comparison to baselines.}
	Finally, We compared our AdaTriplet loss with AutoMargin to competitive metric learning baselines such as SoftTriplet~\cite{qian2019softtriple}, ArcFace~\cite{deng2019arcface}, TL (Triplet)~\cite{hoffer2015deep,schroff2015facenet}, CL (Contrastive)~\cite{chopra2005learning}, WAT~\cite{zhao2019weakly}, and Selectively Contrastive Triplet (SCT)~\cite{xuan2020hard}. Whereas SoftTriplet and ArcFace are classification losses, the other baselines are embedding losses. In~\cref{fig:performances}, our empirical results show that the classification losses generalized poorly on the two test sets, especially on chest X-ray data. On both test sets, our loss outperformed all baselines across years. Notably, on the OAI data, the differences between our method and the baselines were more significant at later years. We present more detailed results in~\Cref{tbl:oai_performances,tbl:cxr_performances}. Moreover, we demonstrate the retrieval results of our method alongside the baselines in~\cref{fig:matched_samples} and~\Cref{fig:matching_samples_more}.

	\section{Discussion}
	In this work, we analyzed Triplet loss in optimizing hard negative samples. To address the issue, we proposed the AdaTriplet loss, whose gradients are adaptive depending on the difficulty of negative samples. In addition, we proposed the AutoMargin method to adjust margin hyperparameters during training. We applied our methodology to the FMIM problem, where the issue of hard negative samples is evident; many medical images may look alike, and it is challenging to capture relevant fine-grained information. Our experiments on two medical datasets showed that AdaTriplet and AutoMargin were robust to visual changes caused by aging and degenerative disorders. The main limitation of this work is that we did not test other neural network architectures, and used grayscale images. However, as recommended in~\cite{musgrave2020metric}, we aimed to make our protocol standard to analyze all the components of the method. Future work should investigate a wider set of models and datasets. We hope our method will be used for other CBMIR tasks, and have made our code publicly available at \url{https://github.com/Oulu-IMEDS/AdaTriplet}.
	
	\section*{Acknowledgments}
	The OAI is a public-private partnership comprised of five contracts (N01- AR-2-2258; N01-AR-2-2259; N01-AR-2- 2260; N01-AR-2-2261; N01-AR-2-2262) funded by the National Institutes of Health, a branch of the Department of Health and Human Services, and conducted by the OAI Study Investigators. Private funding partners include Merck Research Laboratories; Novartis Pharmaceuticals Corporation, GlaxoSmithKline; and Pfizer, Inc. Private sector funding for the OAI is managed by the Foundation for the National Institutes of Health.
	
	We would like to thank the strategic funding of the University of Oulu,  the Academy of Finland Profi6 336449 funding program,  the Northern Ostrobothnia hospital district, Finland (VTR project K33754) and Sigrid Juselius foundation for funding this work. Furthermore, the authors wish to acknowledge CSC – IT Center for Science, Finland, for generous computational resources.
	
	Finally, we thank Matthew B. Blaschko for useful discussions in relation to this paper. Terence McSweeney is acknowledged for proofreading this work and providing comments that improved the clarity of the manuscript. 
	%
	%
	%
	\bibliographystyle{splncs04}
	\bibliography{ref}
	\clearpage
	\newpage

	\setcounter{page}{1}
	\setcounter{figure}{0}
	\setcounter{table}{0}
	\setcounter{section}{0}
	
	\title{AdaTriplet: Adaptive Gradient Triplet Loss -- Supplementary Material}
	
	\titlerunning{AdaTriplet: Adaptive Gradient Triplet Loss -- Supplementary Material}
	%
	
	\author{Anonymous}
	\institute{Anonymous Organization \\
		\email{**@*****.**}}
	
	\author{Khanh Nguyen\thanks{Equal contributions} \and
		Huy Hoang Nguyen$^\star$ \and
		Aleksei Tiulpin}
	%
	
	%
	\authorrunning{Nguyen et al.}
	\institute{University of Oulu, Oulu, Finland \\
		\email{\{khanh.nguyen,huy.nguyen,aleksei.tiulpin\}@oulu.fi}}
	
	\maketitle
	\setcounter{equation}{8}
	
	
	\begin{figure*}[th!]
		\centering
		\IfFileExists{figures/ucircle/ucircle_triplet_loss-crop.pdf}{}{\immediate\write18{pdfcrop 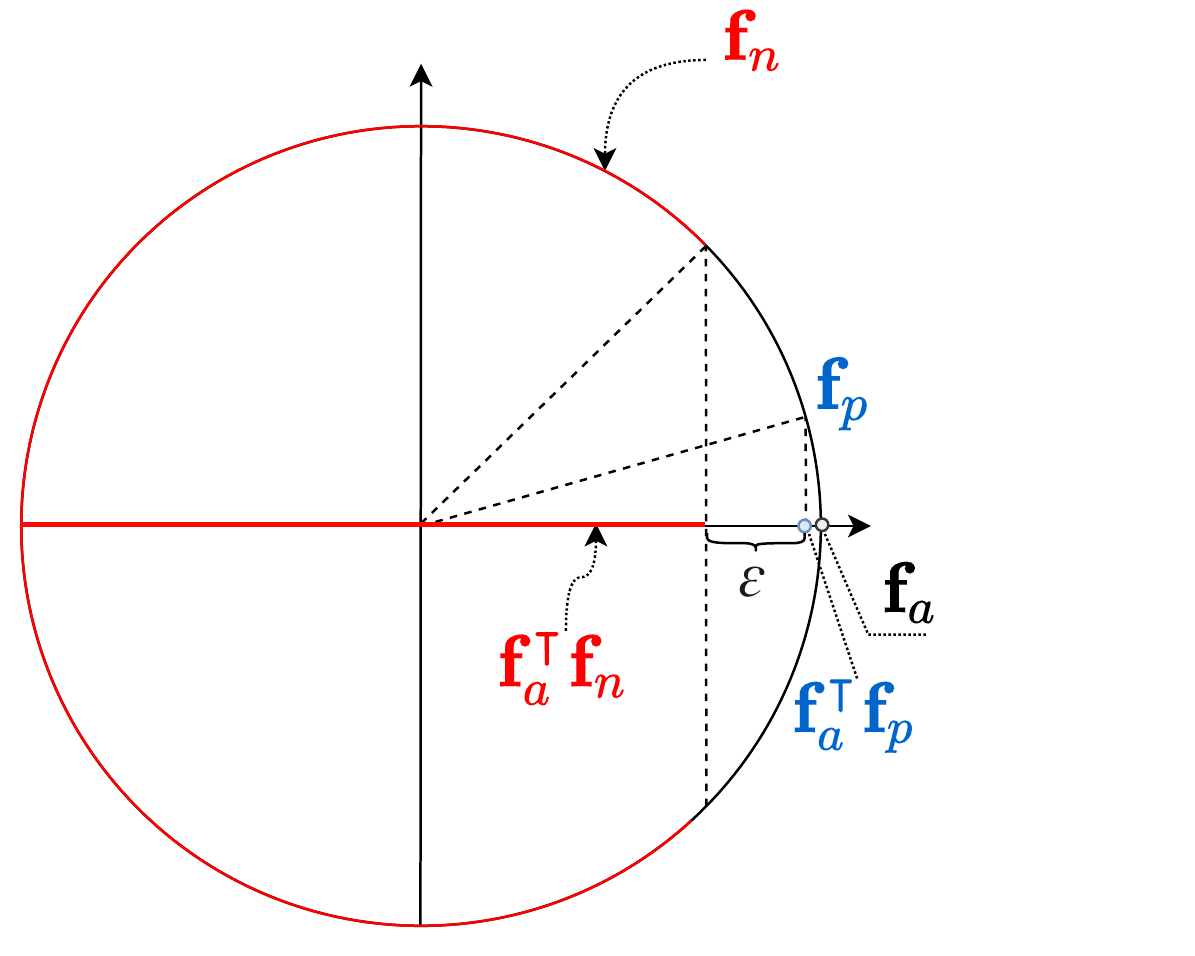}}
		\IfFileExists{figures/ucircle/ucircle_ours1-crop.pdf}{}{\immediate\write18{pdfcrop 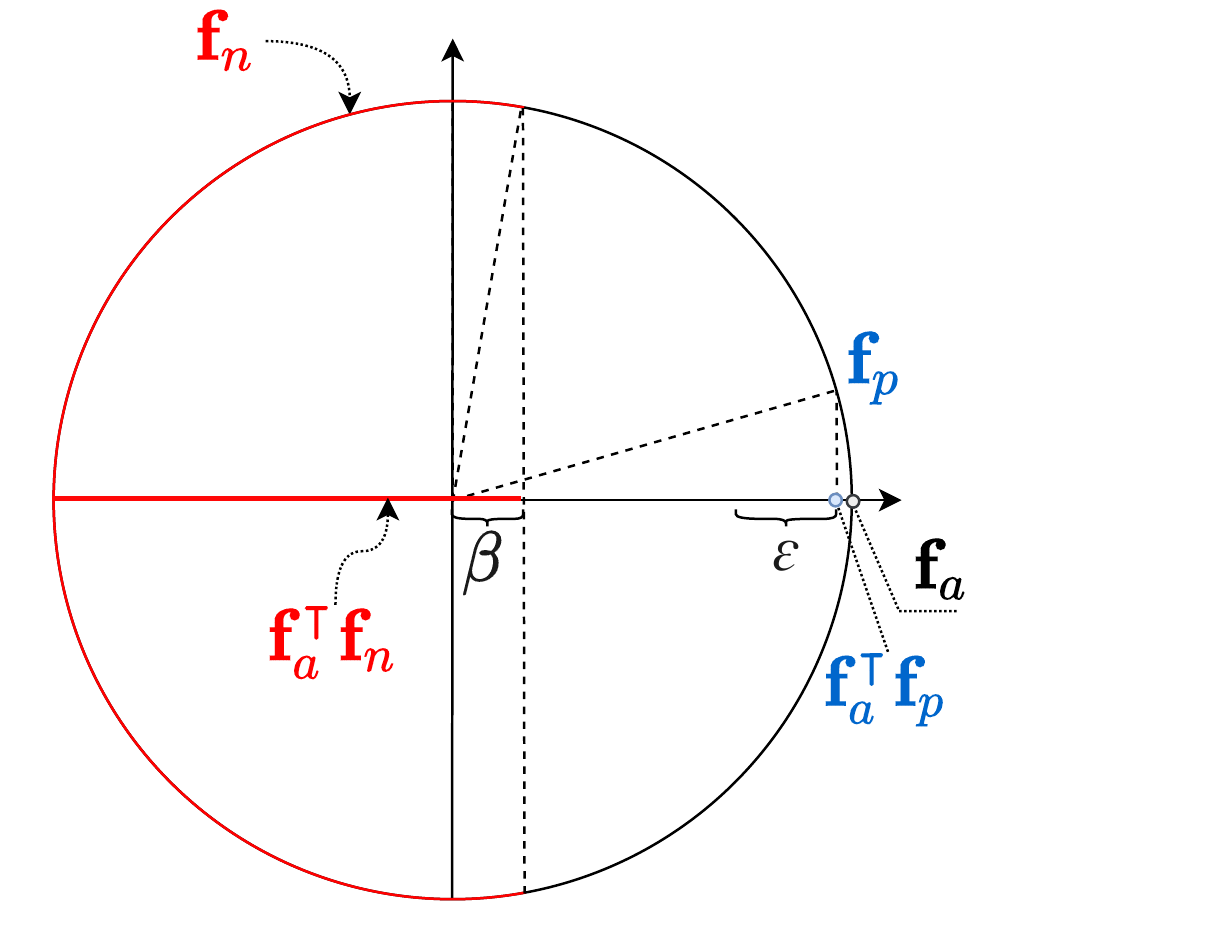}}
		\IfFileExists{figures/ucircle/ucircle_ours2-crop.pdf}{}{\immediate\write18{pdfcrop 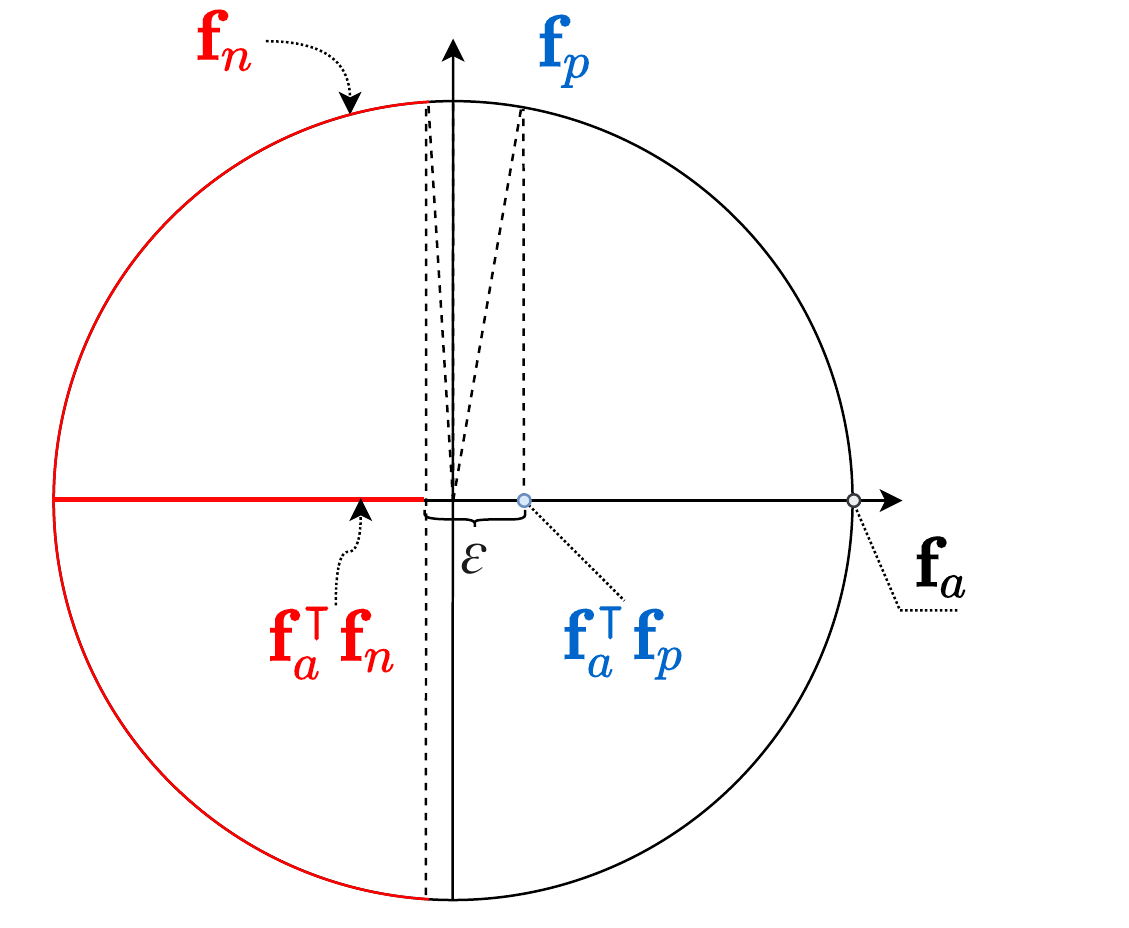}}
		\subfloat[\small Triplet loss ($\bm{f}_a$ and $\bm{f}_p$ are close)][\small Triplet loss\\($\bm{f}_a$ and $\bm{f}_p$ are close)]{\includegraphics[width=0.3\textwidth]{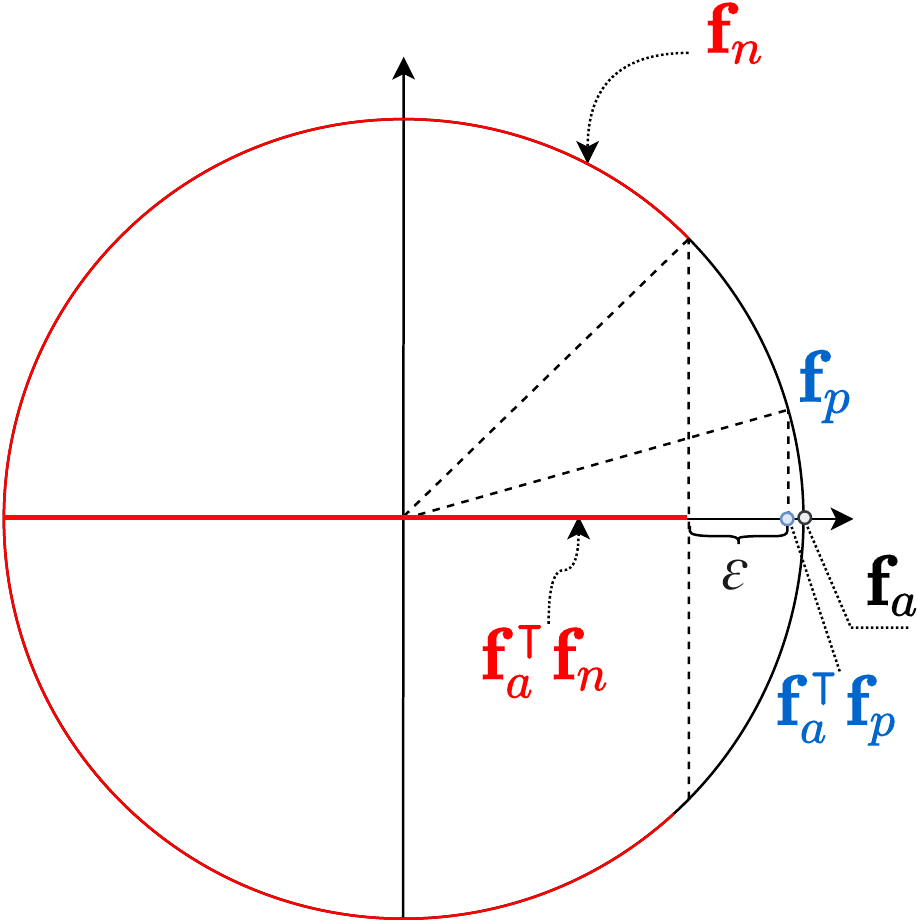}
		}
		\hfill%
		\subfloat[\small AdaTriplet loss ($\bm{f}_a$ and $\bm{f}_p$ are close)][\small AdaTriplet loss\\($\bm{f}_a$ and $\bm{f}_p$ are close)]{\includegraphics[width=0.3\textwidth]{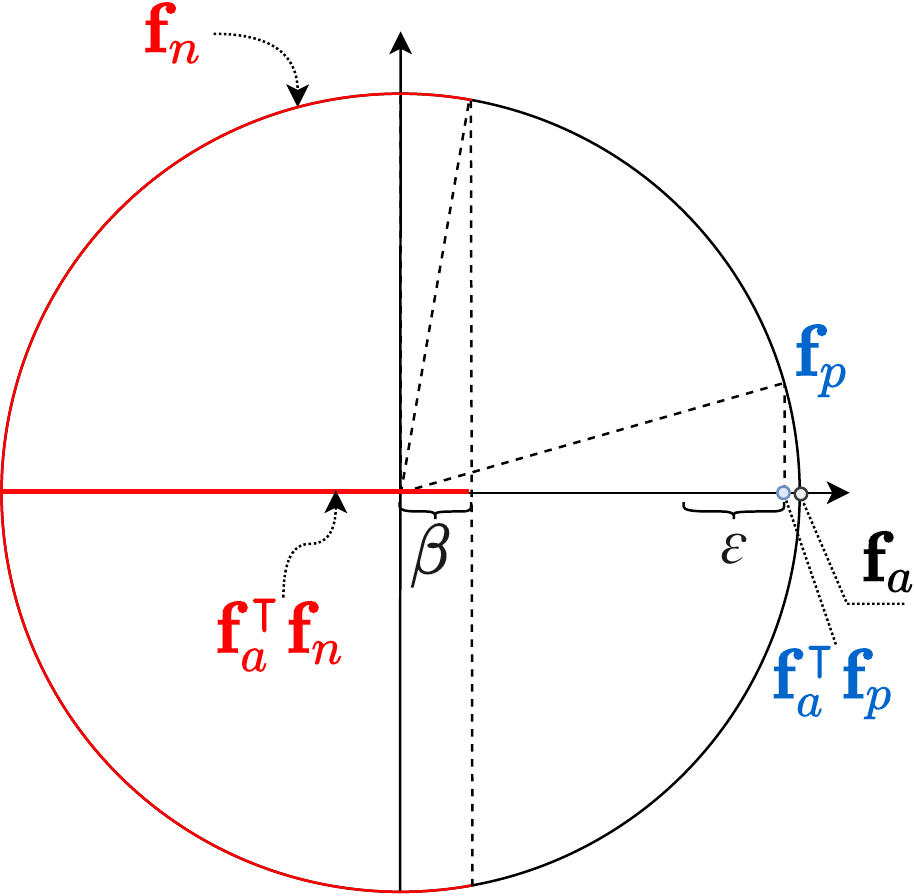}
		}
		\hfill%
		\subfloat[\small AdaTriplet loss ($\bm{f}_a$ and $\bm{f}_p$ are distant)
		][\small AdaTriplet loss\\($\bm{f}_a$ and $\bm{f}_p$ are distant)]{\includegraphics[width=0.3\textwidth]{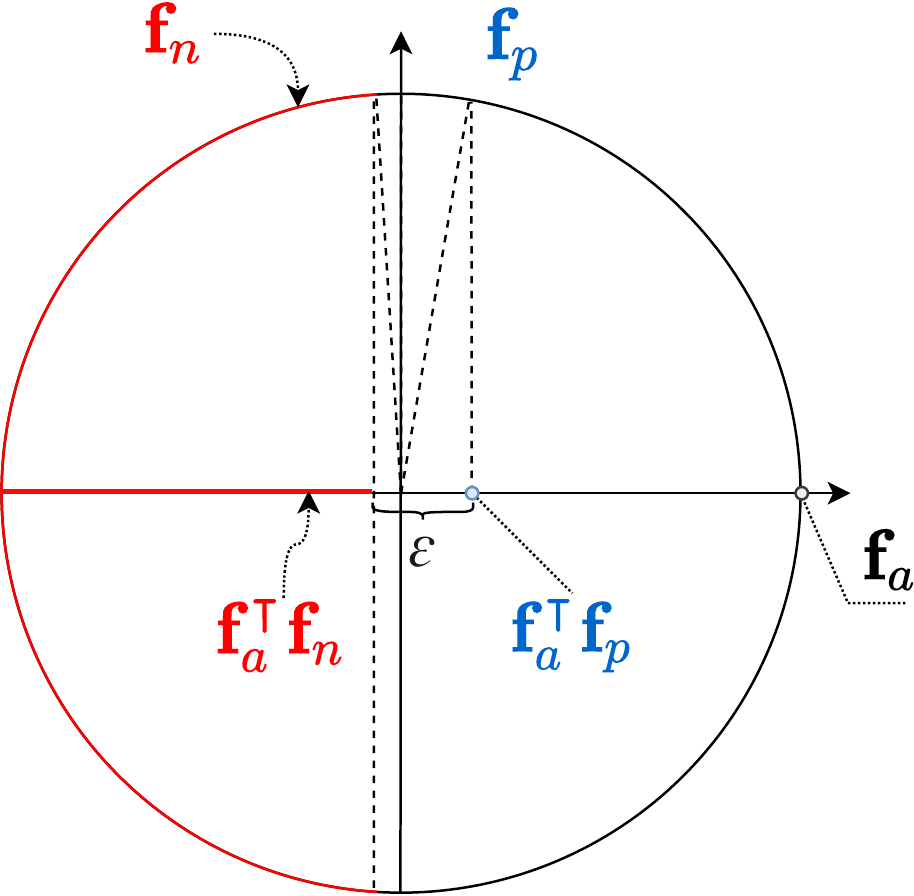}
		}
		\caption{\small Demonstration of triplets of 2D normalized feature vectors on unit circles. Assume that $\bm{f}_a$ is $(1,0)$, then $\bm{f}_a^\intercal\bm{f}_p$ and $\bm{f}_a^\intercal\bm{f}_n$ are the projections of $\bm{f}_p$ and $\bm{f}_n$ on the horizontal axis, respectively. $\varepsilon$ and $\beta$ are margin variables. Red arcs indicate feasible values of $\bm{f}_n$ under a loss function's constraint, and red segments indicate corresponding values of $\bm{f}_a^\intercal\bm{f}_n$. (a) When the angle between $\bm{f}_p$ and $\bm{f}_a$ is small, $\bm{f}_n$ is allowed to be close to $\bm{f}_a$ under the constraint of the Triplet loss~\eqref{eq:triplet_loss_norm}. (b) In the same scenario, our loss has a term to ensure $\bm{f}_n$ to be far from $\bm{f}_a$ at least $\arccos(\beta)$ radian. (c) When $\bm{f}_p$ is sufficiently far from $\bm{f}_a$, the margin $\varepsilon$ overrides the effect of $\beta$.
		}
		\label{fig:ucircle}
	\end{figure*}
	
	\setcounter{proposition}{2}
	
	
	\section{Proof of Proposition 1}
	\label{proof:prop1}
	\begin{proof}
		Consider that for vectors $\bm{a}$ and $\bm{b}$, s.t.  $\|\bm{a}\|_2 = \|\bm{b} \|_2 = 1$. Then $\|\bm{a} - \bm{b}\|^2_2=(\bm{a}-\bm{b})^\intercal(\bm{a}-\bm{b})=2-2\bm{a}^\intercal \bm{b}$. Therefore,
		\begin{align}
		\mathcal{L}_{\mathrm{Triplet}} = \left [2\phi_{an} - 2\phi_{ap}+ \varepsilon \right ]_{+},\ \varepsilon \in [0, 4).
		\end{align}
		By simplifying the coefficient and adjust the range of $\varepsilon$ accordingly, we derive Eq.~\eqref{eq:triplet_loss_norm}.
	\end{proof}
	
	\section{Proof of Proposition 2}
	\label{proof:prop2}
	\begin{proof}
		Let $\mathcal{T}_{+}=\{(\bm{x}_a, \bm{x}_p, \bm{x}_n) \mid \phi_{an} - \phi_{ap} + \varepsilon > 0\}$ and $\mathcal{P}_{+}=\{(\bm{x}_a, \bm{x}_p, \bm{x}_n) \mid \phi_{an} > \beta \}$ denote the set of all not-easy triplets and the set of all triplets with a hard negative pair, respectively. Then, the AdaTriplet loss intrinsically partitions the domain of the loss function in into $4$ sub-domains:
		\begin{align}
		\mathcal{L}_{\mathrm{AdaTriplet}}(\tau) = \left\{\begin{matrix}
		(1+\lambda)\phi_{an} - \phi_{ap} + \varepsilon - \lambda\beta& \mathrm{if}\ \tau \in \mathcal{T}_{+} \cap \mathcal{P}_{+} \\ 
		\lambda \phi_{an} - \lambda\beta & \mathrm{if}\ \tau \in (\mathcal{T} \textbackslash \mathcal{T}_{+}) \cap \mathcal{P}_{+} \\ 
		\phi_{an} - \phi_{ap} +\varepsilon & \mathrm{if}\ \tau \in \mathcal{T}_{+} \cap (\mathcal{T} \textbackslash \mathcal{P}_{+}) \\ 
		0 & \mathrm{otherwise},
		\end{matrix}\right. \ ,
		\end{align}
		where $\lambda \in \mathbb{R}_{+}$, and $\tau \in \mathcal{T}$ is a triplet. As a result, we can derive the partial derivatives of $\mathcal{L}_{\mathrm{AdaTriplet}}$ with respect to $\phi_{ap}$ and $\phi_{an}$
		\begin{align}
		\left (\frac{\partial \mathcal{L}_{\mathrm{AdaTriplet}}(\tau)}{\partial \phi_{ap}} , \frac{\partial \mathcal{L}_{\mathrm{AdaTriplet}}(\tau)}{\partial \phi_{an}}\right) = \left\{\begin{matrix}
		(-1, 1+\lambda)& \mathrm{if}\ \tau \in \mathcal{T}_{+} \cap  \mathcal{P}_{+}  \\ 
		(0, \lambda) & \mathrm{if}\ \tau \in (\mathcal{T} \textbackslash \mathcal{T}_{+}) \cap  \mathcal{P}_{+}  \\
		(-1, 1) & \mathrm{if}\ \tau \in \mathcal{T}_{+} \cap  (\mathcal{T} \textbackslash \mathcal{P}_{+})  \\ 
		(0, 0) & \mathrm{otherwise}.
		\end{matrix}\right.
		\end{align}
		
		In contrast, $\left (\frac{\partial \mathcal{L}^{*}_{\mathrm{Triplet}}(\tau)}{\partial \phi_{ap}} , \frac{\partial \mathcal{L}^{*}_{\mathrm{Triplet}}(\tau)}{\partial \phi_{an}}\right) = (-1, 1), \forall \tau \in \mathcal{T}_{+}$, which concludes the proof.
	\end{proof}
	
	\begin{table}[hb!]
		\centering
		\caption{\small (a) An ordered list of common transformations. ($\checkmark$) indicates ones only used in the training phase. (b) Lists of hyperparameter values. Bold and underlined numbers indicate selected values for OAI and CXR, respectively.}
		\renewcommand{\arraystretch}{1.2}
		\subfloat[\label{tbl:oai_train_augmentation}]{
			\scalebox{0.75}{
				\begin{tabular}{lcr}
					\toprule
					\multicolumn{1}{l}{\textBF{Transformation}} & \textBF{Prob.} & \textBF{Parameter} \\
					\midrule 
					Resize & 1 & $280 \times 280$ \\
					Gaussian noise ($\checkmark$) & 0.5   & 0.3 \\
					Rotation ($\checkmark$) & 1     & [-10, 10] \\
					Random cropping ($\checkmark$) & 1     & $256\times256$ \\
					Center cropping & 1     & $256\times256$ \\
					Gamma correction ($\checkmark$) & 0.5   & [0.5, 1.5] \\
					\multirow{2}{*}{Normalization} & \multirow{2}{*}{1} & [0.5, 0.3] on OAI\\
					& & [0.5, 0.5] on CXR \\
					\bottomrule
			\end{tabular}}
			
		}\hfill
		\subfloat[\label{tbl:hyperparameter_range}]{
			\scalebox{0.75}{
				\begin{tabular}{lcr}
					\toprule
					\textBF{Method}      & \textBF{Hyperparam.} & \textBF{List of values}                    \\
					\midrule 
					SCT         & $\lambda$       & \{0, \hlCXR{0.5}, \hlOAI{1}, 2\}            \\ \midrule
					ArcFace     & $m$           & \{5.7, \hlOAI{14.3}, \hlCXR{28.6}, 43\}     \\ \midrule
					WAT         & $\beta$        & \{\hlCXR{\hlOAI{0.1}}, 0.25, 0.5, 0.75\} \\ \midrule
					SoftTriplet & $m$           & \{\hlCXR{0.01}, \hlOAI{0.02}, 0.05, 0.1\}   \\ \midrule
					\multirow{2}{*}{Contrastive} & $m_{neg}$     & \{0.25, 0.5, \hlCXR{\hlOAI{0.75}}, 1\}   \\
					& $m_{pos}$      & \{0, 0.25, \hlCXR{\hlOAI{0.5}}, 0.75\}   \\ 
					\midrule
					Triplet     &  & \\ 
					+ Grid search  & $\varepsilon$    & \{0.1, \hlOAI{0.25}, \hlCXR{0.5}, 0.75\} \\
					\cmidrule{2-3}
					+ AutoMargin  & $K_\Delta$    & \{\hlCXR{\hlOAI{2}}, 4\}                  \\
					\midrule
					AdaTriplet & & \\
					\multirow{2}{*}{+ Grid search}  & $\varepsilon$    & \{0.1, 0.25, \hlCXR{\hlOAI{0.5}}, 0.75\} \\
					& $\beta$          & \{0.1, 0.25, \hlOAI{0.5}, \hlCXR{0.75}\} \\
					\cmidrule{2-3}
					\multirow{2}{*}{+ AutoMargin}  & $K_\Delta$    & \{\hlCXR{\hlOAI{2}}, 4\}                  \\
					& $K_{an}$      & \{\hlOAI{2}, \hlCXR{4}\}                  \\
					\bottomrule
			\end{tabular}}
			
		}
	\end{table}
	
	\begin{table}[hb!]
		\caption{\small Descriptions of the OAI and CXR datasets. Knee X-ray images with disease indicate those with KL grade greater than $1$. Chest X-ray images with disease consist of those with at least one lung or heart disease. OAI test data are from the acquisition site C. The CXR data splits are given by the CXR's owner. Both the galleries contain only data points at their baselines. Queries are from the other follow-up visits.}
		\centering
		\addtolength{\tabcolsep}{5pt}
		\renewcommand{\arraystretch}{1.2}
		\subfloat[Overview descriptions]{
			\scalebox{0.8}{
				\begin{tabular}{clccccr}
					\toprule
					\textBF{Dataset} & \textBF{Phase} & \textBF{\# Images}            & \textBF{\# Subjects}      & \textBF{\% Male}           & \textBF{\# Images with disease} \\
					\midrule
					\multirow{2}{*}{OAI}   &  Training/validation &       37410 &   3490 &  59.1 &   14240 \\
					&   Test  &  15648 &   1306 &   53.8 &   5409  \\
					\midrule
					\multirow{2}{*}{CXR}        &    Training/validation  &   86524 &   28008 &   56.0 &   36324 \\
					& Test		& 25587 &       2797 &   58.1 &   15735 \\
					\bottomrule
				\end{tabular} 
		}}
		\\
		\subfloat[\small Detailed descriptions of the test sets.]{
			\scalebox{0.8}{
				\begin{tabular}{lrrrrrrrrrrrrr}
					\toprule
					& & \multicolumn{12}{c}{\textBF{Query (at year)}} \\
					\cmidrule{3-14}
					\textBF{Dataset}           & \textBF{Gallery} & \textBF{1}            & \textBF{2}           & \textBF{3}           & \textBF{4}    & \textBF{5}       & \textBF{6}           & \textBF{7}   & \textBF{8}    & \textBF{9}  & \textBF{10}  & \textBF{11} & \textBF{12}           \\ \midrule
					{OAI}  &   2610  & 2498 & 2430 & 2306 & 2252 & 0   & 1858 & 0   & 1694 & 0   & 0   & 0  & 0  \\
					{CXR}  & 13137 & 5662 & 1944 & 1216 & 1123 & 636 & 528  & 541 & 236  & 304 & 105 & 75 & 80 \\

					\bottomrule
				\end{tabular}
			}
		}
		\label{tbl:data_descriptions}
	\end{table}
	
	\begin{table}[hp!]
		\caption{\small Comparison between an exhaustive grid search for fixed margins and AutoMargin for adaptive margins in the AdaTriplet loss on OAI and CXR. SE means standard error.}
		\centering
		\addtolength{\tabcolsep}{5pt}
		\hspace*{\fill}
		
		\subfloat[Exhaustive grid search\label{tbl:exhaustive_grid_search}]{
			\scalebox{0.8}{
				\begin{tabular}{cccc}
					\toprule
					\textBF{$\varepsilon$}            & \textBF{$\beta$} & \textBF{mAP$_{OAI}$ (\%)} &  \textBF{mAP$_{CXR}$ (\%)}       \\
					\midrule
					&         0.1 &      95.90 & 79.33 \\
					&       0.25 &      96.15 & 80.08\\
					&        0.5 &      96.27 & 79.50\\
					\multirow{-4}{*}{0.1}  &       0.75 &      96.20 & 80.81 \\
					\midrule
					&        0.1 &      96.20 & 83.48\\
					&       0.25 &      96.66 & 84.15\\
					&        0.5 &      96.91 & 84.40\\
					\multirow{-4}{*}{0.25}     &       0.75 &      96.75 & 85.27 \\
					\midrule
					&        0.1 &      95.70 & 84.07\\
					&       0.25 &      96.36 & 86.04\\
					&        0.5 &      \textBF{97.02} & 85.84\\
					\multirow{-4}{*}{0.5}            &       0.75 &      96.44 & \textBF{86.65} \\
					\midrule
					&        0.1 &      92.10 & 79.29\\
					&       0.25 &      95.59 & 85.42\\
					&        0.5 &      96.59 & 81.75\\
					\multirow{-4}{*}{0.75}  &       0.75 &      95.33 & 82.89\\
					\midrule
					\multicolumn{2}{l}{Mean$\pm$SE} & 96.01$\pm$0.28 & 83.06$\pm$0.65\\
					\bottomrule
			\end{tabular} }
		}
		\hfill
		\subfloat[AutoMargin\label{tbl:AdaTriplet_grid_search}]{
			\scalebox{0.8}{
				\begin{tabular}{cccc}
					\toprule
					
					\textBF{$K_\Delta$}            & \textBF{$K_{an}$} & \textBF{mAP$_{OAI}$ (\%)} &  \textBF{mAP$_{CXR}$ (\%)}      \\
					\midrule
					&          2 &     \textBF{97.08} & 85.95\\
					\multirow{-2}{*}{2}  &          4 &     96.70 & \textBF{87.04} \\
					\midrule
					&          2 &     96.58 & 83.71\\
					\multirow{-2}{*}{4}      &          4 &     96.73 & 85.28 \\
					\midrule
					\multicolumn{2}{l}{Mean$\pm$SE} & 96.77$\pm$0.09 & 85.50$\pm$0.70\\
					\bottomrule
				\end{tabular}
		}}
		\hspace*{\fill}
		\label{tbl:grid_search}
	\end{table}

	\begin{table*}[t!]
		\caption{\small Performance comparisons on the OAI test set (mean and standard error over $5$ random seeds). Bold values indicate the best performances, and underline values indicate ones that are substantially higher than the others. Rows corresponding to our method are highlighted. }
		\centering
		\addtolength{\tabcolsep}{5pt}
		\resizebox{1.\textwidth}{!}{
			\begin{tabular}{llkhhhhhhh}
				\toprule
				\rowcolor{white}
				\multicolumn{2}{c}{\textBF{Metric}}            & \textBF{Loss} & \textBF{1 year}            & \textBF{2 years}           & \textBF{3 years}           & \textBF{4 years}           & \textBF{6 years}           & \textBF{8 years}           & \textBF{All}           \\ \midrule
				\rowcolor{white}
				&           &  SoftTriplet &         92.1$\pm$0.2 &   91.2$\pm$0.3 &   85.8$\pm$0.6 &   79.4$\pm$1.1 &   73.3$\pm$1.2 &   71.3$\pm$1.4 &   83.2$\pm$0.7 \\
				
				\rowcolor{white}
				
				&							& ArcFace &        93.0$\pm$0.1 &   92.0$\pm$0.1 &   86.5$\pm$0.2 &   80.0$\pm$0.7 &   74.7$\pm$1.0 &   72.7$\pm$0.8 &   84.2$\pm$0.4 \\
				
				\rowcolor{white} 
				&							& SCT &          97.1$\pm$0.1 &   96.6$\pm$0.1 &   94.4$\pm$0.2 &   87.5$\pm$0.7 &   83.4$\pm$0.9 &   81.6$\pm$0.9 &   90.9$\pm$0.4 \\
				\rowcolor{white} 
				&							& WAT &            98.1$\pm$0.0 &   97.7$\pm$0.0 &   96.6$\pm$0.1 &   92.0$\pm$0.6 &   89.4$\pm$0.8 &   88.1$\pm$0.9 &   94.2$\pm$0.3 \\
				
				\rowcolor{white}
				&		& Contrastive &             98.3$\pm$0.0 &   97.9$\pm$0.0 &   97.1$\pm$0.0 &   93.4$\pm$0.2 &   91.1$\pm$0.2 &   90.2$\pm$0.2 &   95.1$\pm$0.1 \\

				\rowcolor{white}
				&   & {Triplet}  &       98.1$\pm$0.1 &   97.8$\pm$0.0 &   96.7$\pm$0.1 &   92.4$\pm$0.5 &   89.6$\pm$0.8 &   88.6$\pm$0.8 &   94.4$\pm$0.3 \\

				

				\multicolumn{2}{c}{\multirow{-7}{*}{mAP}}             &  AdaTriplet  &         \subbest{98.5$\pm$0.0} &   \subbest{98.3$\pm$0.0} &   \subbest{97.9$\pm$0.0} &   \subbest{96.2$\pm$0.1} &   \subbest{95.0$\pm$0.2} &   \subbest{94.2$\pm$0.3} &   \subbest{96.9$\pm$0.1} \\
				
				\midrule
				\rowcolor{white}
				&           &  SoftTriplet &           78.2$\pm$0.3 &   75.3$\pm$0.5 &   58.7$\pm$1.3 &   47.7$\pm$1.7 &   37.8$\pm$1.8 &   35.2$\pm$1.9 &   57.6$\pm$1.2 \\
				
				\rowcolor{white}
				
				&							& ArcFace &          83.1$\pm$0.2 &   80.6$\pm$0.3 &   66.5$\pm$0.6 &   56.0$\pm$1.1 &   47.9$\pm$1.2 &   45.3$\pm$1.1 &   65.1$\pm$0.6 \\
				
				\rowcolor{white} 
				&							& SCT &            93.4$\pm$0.3 &   91.8$\pm$0.2 &   85.0$\pm$0.8 &     69.3$\pm$1.0 &   58.3$\pm$1.5 &   56.4$\pm$1.9 &   77.6$\pm$0.8 \\
				
				\rowcolor{white} 
				&							& WAT &              95.7$\pm$0.1 &   95.2$\pm$0.1 &   91.4$\pm$0.2 &   79.0$\pm$1.1 &   70.4$\pm$1.7 &   69.0$\pm$1.9 &   84.9$\pm$0.7 \\
				
				\rowcolor{white}
				&		& Contrastive &               96.4$\pm$0.1 &   95.7$\pm$0.1 &   93.0$\pm$0.1 &   80.9$\pm$0.7 &   72.5$\pm$0.7 &   71.3$\pm$0.8 &   86.3$\pm$0.3 \\

				\rowcolor{white}
				&   & {Triplet}  &         95.8$\pm$0.2 &   94.9$\pm$0.2 &   91.7$\pm$0.3 &   78.4$\pm$1.6 &   68.8$\pm$2.6 &   67.7$\pm$2.5 &   84.4$\pm$1.1 \\



				\multicolumn{2}{c}{\multirow{-7}{*}{mAP$@$R}}             &  AdaTriplet  &         \subbest{97.0$\pm$0.1} &   \subbest{96.3$\pm$0.1} &   \subbest{94.5$\pm$0.2} &   \subbest{87.9$\pm$0.5} &   \subbest{83.9$\pm$0.7} &   \subbest{82.3$\pm$0.8} &   \subbest{91.1$\pm$0.3} \\

				\midrule
				\rowcolor{white}
				&  	& SoftTriplet  &        89.5$\pm$0.2 &   88.2$\pm$0.3 &   80.9$\pm$0.9 &   73.1$\pm$1.3 &   65.4$\pm$1.4 &   63.2$\pm$1.6 &   78.1$\pm$0.8 \\
				
				\rowcolor{white}
				&							& ArcFace &      90.6$\pm$0.1 &   89.2$\pm$0.1 &   81.8$\pm$0.3 &   74.0$\pm$0.9 &   67.3$\pm$1.2 &   64.7$\pm$1.0 &   79.2$\pm$0.5 \\
				\rowcolor{white}
				&		& SCT &            95.6$\pm$0.1 &   94.7$\pm$0.1 &   91.1$\pm$0.3 &     81.4$\pm$1.0 &   75.3$\pm$1.2 &   73.3$\pm$1.2 &   86.4$\pm$0.6 \\
				
				\rowcolor{white}
				&		& WAT &              97.3$\pm$0.1 &   96.7$\pm$0.1 &   94.8$\pm$0.1 &   87.9$\pm$0.8 &   83.8$\pm$1.1 &   82.4$\pm$1.2 &   91.3$\pm$0.5 \\
				
				\rowcolor{white}
				&		& Contrastive &              97.6$\pm$0.0 &   97$\pm$0.1.0 &   95.6$\pm$0.1 &   89.9$\pm$0.3 &   86.2$\pm$0.2 &   85.0$\pm$0.3 &   92.5$\pm$0.1 \\
				
				\rowcolor{white}
				&            & {Triplet}  &        97.3$\pm$0.1 &   96.8$\pm$0.1 &   95.0$\pm$0.2 &   88.6$\pm$0.7 &   84.3$\pm$1.1 &   83.1$\pm$1.1 &   91.6$\pm$0.5 \\
				
				

				\multirow{-7}{*}{CMC}			&	 \multirow{-7}{*}{top 1}		& AdaTriplet  &       \subbest{98.0$\pm$0.1} &   \subbest{97.7$\pm$0.1} &   \subbest{96.9$\pm$0.1} &   \subbest{94.5$\pm$0.2} &   \subbest{92.8$\pm$0.3} &   \subbest{91.6$\pm$0.4} &   \subbest{95.6$\pm$0.2} \\
				\bottomrule
		\end{tabular} }
		\label{tbl:oai_performances}
	\end{table*}
	
	
	\begin{table}
		\caption{\small Performance comparisons on the CXR test set (mean and standard error over $5$ random seeds). Results of our AdaTriplet loss are highlighted. The best performances are in bold, and underline values indicate ones that are substantially higher than the others.}
		\resizebox{\textwidth}{!}{
			\begin{tabular}{llkhhhhhhhhhhhhh}
				\toprule
				\rowcolor{white}
				\multicolumn{2}{c}{\textBF{Metric}}            & \textBF{Loss} & \textBF{1 year}            & \textBF{2 years}           & \textBF{3 years}           & \textBF{4 years}    & \textBF{5 years}       & \textBF{6 years}           & \textBF{7 years}   & \textBF{8 years}    & \textBF{9 years}  & \textBF{10 years}  & \textBF{11 years} & \textBF{12 years} & \textBF{All}           \\ \midrule
				\rowcolor{white}
				&      & {SoftTriplet}  &             27.3$\pm$0.1 &   21.6$\pm$0.3 &   19.3$\pm$0.3 &   17.4$\pm$0.4 &   17.5$\pm$0.3 &   18.3$\pm$0.3 &   10.6$\pm$0.3 &   15.3$\pm$0.8 &   18.4$\pm$2.1 &   12.6$\pm$0.6 &   13.6$\pm$0.6 &    6.5$\pm$0.6 &   22.3$\pm$0.1 \\
				
				\rowcolor{white}
				&      & {ArcFace}  &           29.9$\pm$0.2 &   24.2$\pm$0.1 &   22.1$\pm$0.3 &   18.7$\pm$0.3 &   19.4$\pm$0.3 &   21.2$\pm$0.7 &   12.6$\pm$0.6 &   17.0$\pm$0.7 &   23.0$\pm$1.0 &   14.0$\pm$0.6 &   14.6$\pm$1.3 &    7.4$\pm$0.7 &   24.8$\pm$0.2 \\
				
				\rowcolor{white}
				&      & {SCT}  &             70.2$\pm$0.6 &   61.5$\pm$0.6 &   59.0$\pm$0.3 &   57.3$\pm$0.5 &   54.0$\pm$0.7 &   55.2$\pm$0.6 &   34.9$\pm$0.5 &   57.7$\pm$1.2 &   56.1$\pm$1.7 &   50.4$\pm$1.1 &     62.5$\pm$1.0 &   44.2$\pm$3.0 &   62.6$\pm$0.5 \\
				
				\rowcolor{white}
				&      & {WAT}  &               81.5$\pm$0.4 &   72.7$\pm$0.3 &   66.1$\pm$0.4 &   67.1$\pm$0.4 &   64.2$\pm$0.6 &   67.1$\pm$1.2 &   43.1$\pm$0.5 &   65.8$\pm$0.4 &   65.6$\pm$0.8 &   69.9$\pm$0.6 &   74.7$\pm$0.8 &   57.9$\pm$2.4 &   73.2$\pm$0.3 \\
				
				\rowcolor{white}
				&      & {Contrastive}  &   79.7$\pm$0.4 &   71.4$\pm$0.6 &   65.5$\pm$0.2 &   66.3$\pm$0.3 &   64.6$\pm$0.3 &   66.1$\pm$1.1 &   43.7$\pm$0.5 &   65.7$\pm$1.2 &   65.2$\pm$1.5 &   69.7$\pm$1.4 &   72.7$\pm$0.7 &   52.3$\pm$3.0 &   72.0$\pm$0.4 \\

				\rowcolor{white}
				&      & {Triplet}  &         80.9$\pm$0.6 &   71.3$\pm$0.4 &   65.3$\pm$0.6 &   65.9$\pm$0.3 &   64.4$\pm$0.3 &   67.7$\pm$0.5 &   42.2$\pm$0.3 &   66.2$\pm$1.6 &   66.2$\pm$0.4 &   66.8$\pm$1.9 &   73.2$\pm$0.3 &   55.4$\pm$4.3 &   72.5$\pm$0.4 \\
				

				\multicolumn{2}{c}{\multirow{-7}{*}{mAP}}             &  AdaTriplet &    \subbest{82.5$\pm$0.5} &   \subbest{74.5$\pm$0.2} &   \unsubbest{67.0$\pm$0.7} &   \subbest{68.0$\pm$0.5} &   \subbest{65.6$\pm$0.6} &   \unsubbest{68.7$\pm$1.1} &   \unsubbest{44.0$\pm$0.7} &   \subbest{68.8$\pm$1.0} &   \subbest{71.1$\pm$1.6} &   \unsubbest{72.3$\pm$1.7} &   \unsubbest{74.8$\pm$1.1} &   \unsubbest{59.6$\pm$1.3} &   \subbest{74.5$\pm$0.4} \\

				\midrule
				\rowcolor{white}
				&      & {SoftTriplet}  &    14.9$\pm$0.1 &   11.0$\pm$0.2 &    8.8$\pm$0.3 &    7.9$\pm$0.4 &    8.2$\pm$0.2 &    8.2$\pm$0.4 &    4.6$\pm$0.3 &    5.8$\pm$0.8 &    9.1$\pm$1.3 &    3.6$\pm$0.4 &    4.2$\pm$0.9 &    1.3$\pm$0.5 &   11.4$\pm$0.1 \\
				
				\rowcolor{white}
				&      & {ArcFace}  &    16.7$\pm$0.1 &   12.2$\pm$0.3 &   10.4$\pm$0.3 &    9.0$\pm$0.5 &    9.0$\pm$0.6 &    9.5$\pm$0.4 &    5.5$\pm$0.5 &    7.0$\pm$0.3 &    9.5$\pm$1.6 &    5.1$\pm$0.8 &    5.2$\pm$0.5 &    1.4$\pm$0.3 &   12.9$\pm$0.2 \\
				
				\rowcolor{white}
				&      & {SCT}  &         57.3$\pm$0.6 &   50.3$\pm$0.9 &   47.5$\pm$0.3 &   46.0$\pm$0.9 &   42.6$\pm$0.6 &   43.9$\pm$0.8 &   26.9$\pm$0.5 &   43.1$\pm$1.4 &   36.7$\pm$2.8 &   34.0$\pm$2.1 &   46.9$\pm$0.9 &   28.9$\pm$2.2 &   50.4$\pm$0.6 \\
				
				\rowcolor{white}
				&      & {WAT}  &       71.1$\pm$0.7 &   62.5$\pm$1.0 &   56.5$\pm$0.4 &   58.2$\pm$0.5 &   54.3$\pm$0.5 &   56.9$\pm$1.0 &   35.0$\pm$0.6 &   55.0$\pm$0.8 &   50.1$\pm$1.9 &   56.3$\pm$2.1 &   67.0$\pm$0.5 &   \unsubbest{47.0$\pm$1.9} &   63.0$\pm$0.6 \\
				
				\rowcolor{white}
				&      & {Contrastive}  &     69.4$\pm$0.6 &   61.5$\pm$0.9 &   56.3$\pm$0.4 &   56.8$\pm$0.7 &   55.9$\pm$0.5 &   57.5$\pm$0.7 &   36.5$\pm$0.5 &   56.8$\pm$1.7 &   48.1$\pm$1.3 &   57.5$\pm$2.1 &   64.1$\pm$1.7 &   43.9$\pm$4.0 &   62.1$\pm$0.5 \\
				
				\rowcolor{white}
				&      & {Triplet}  &     70.2$\pm$0.8 &   61.5$\pm$0.8 &   56.2$\pm$0.7 &   55.7$\pm$0.6 &   54.1$\pm$0.8 &   58.2$\pm$0.8 &   33.4$\pm$0.5 &   54.6$\pm$1.1 &   48.9$\pm$0.3 &   54.4$\pm$2.0 &   64.9$\pm$1.4 &   43.4$\pm$2.8 &   62.1$\pm$0.6 \\
				
				
				
				\multicolumn{2}{c}{\multirow{-7}{*}{mAP$@$R}}             &  AdaTriplet &      \subbest{72.3$\pm$0.7} &   \subbest{64.9$\pm$0.6} &   \subbest{58.2$\pm$1.1} &     \unsubbest{58.9$\pm$1.0} &   \unsubbest{56.4$\pm$0.7} &   \unsubbest{59.6$\pm$1.6} &   \unsubbest{37.7$\pm$0.9} &   \unsubbest{59.1$\pm$1.6} &   \subbest{58.6$\pm$1.9} &   \subbest{62.8$\pm$2.6} &   \unsubbest{68.8$\pm$2.6} &   46.7$\pm$1.8 &   \subbest{64.9$\pm$0.5} \\

				\midrule
				\rowcolor{white}
				&  & {SoftTriplet}  &        20.3$\pm$0.2 &   16.5$\pm$0.3 &   14.6$\pm$0.3 &   14.0$\pm$0.3 &   13.7$\pm$0.3 &   13.7$\pm$0.4 &    8.20$\pm$0.2 &   11.5$\pm$0.7 &   14.8$\pm$2.2 &    9.30$\pm$0.4 &    9.80$\pm$0.9 &    4.40$\pm$0.8 &   16.9$\pm$0.2 \\
				
				\rowcolor{white}
				&  & {ArcFace}  &      22.9$\pm$0.2 &   18.9$\pm$0.2 &   17.2$\pm$0.3 &   15.3$\pm$0.3 &   15.3$\pm$0.3 &   16.3$\pm$0.5 &   10.0$\pm$0.6 &   13.5$\pm$0.5 &   18.5$\pm$1.0 &   11.1$\pm$0.8 &   11.0$\pm$1.3 &    5.4$\pm$0.8 &   19.2$\pm$0.1 \\
				
				\rowcolor{white}
				&  & {SCT}  &        57.2$\pm$0.5 &   50.3$\pm$0.6 &   50.2$\pm$0.3 &   48.5$\pm$0.6 &   44.5$\pm$0.8 &   46.3$\pm$0.6 &   29.2$\pm$0.4 &   48.1$\pm$1.3 &   45.9$\pm$2.2 &   39.6$\pm$1.0 &   51.1$\pm$1.4 &   34.0$\pm$2.9 &   51.6$\pm$0.4 \\
				
				\rowcolor{white}
				&  & {WAT}  &          71.4$\pm$0.5 &   64.1$\pm$0.5 &   58.9$\pm$0.4 &   59.9$\pm$0.4 &   56.1$\pm$0.6 &   58.0$\pm$0.9 &   37.2$\pm$0.5 &   58.5$\pm$0.4 &   56.2$\pm$1.2 &   60.4$\pm$0.6 &   68.2$\pm$0.8 &   47.4$\pm$2.7 &   64.3$\pm$0.4 \\
				
				\rowcolor{white}
				&      & {Contrastive}  &     69.8$\pm$0.5 &   62.5$\pm$0.8 &   58.2$\pm$0.3 &   59.3$\pm$0.3 &   57.5$\pm$0.5 &   57.9$\pm$1.2 &   38.0$\pm$0.4 &   58.3$\pm$1.5 &   56.3$\pm$1.8 &   61.3$\pm$1.1 &   66.4$\pm$1.0 &   43.4$\pm$2.7 &   63.3$\pm$0.5 \\
				
				\rowcolor{white}
				&  & {Triplet}  &     70.5$\pm$0.8 &   62.5$\pm$0.6 &   57.7$\pm$0.7 &   58.5$\pm$0.3 &   56.0$\pm$0.4 &   58.4$\pm$0.6 &   36.0$\pm$0.2 &   58.1$\pm$1.5 &   56.1$\pm$1.9 &   57.9$\pm$0.4 &   65.4$\pm$0.8 &   46.4$\pm$3.6 &   63.3$\pm$0.6 \\
				

				\multirow{-7}{*}{CMC}		&	\multirow{-7}{*}{top 1} 		& AdaTriplet  &       \subbest{72.9$\pm$0.6} &   \subbest{66.1$\pm$0.3} &   \unsubbest{60.0$\pm$0.8} &   \subbest{61.0$\pm$0.6} &   \unsubbest{58.1$\pm$0.6} &   \subbest{60.3$\pm$0.9} &   \unsubbest{38.3$\pm$0.8} &   \subbest{61.3$\pm$0.8} &   \subbest{62.6$\pm$2.0} &   \subbest{63.7$\pm$1.8} &   \unsubbest{68.5$\pm$1.7}&   \unsubbest{47.7$\pm$0.9} &   \subbest{66.0$\pm$0.4} \\

				\bottomrule
			\end{tabular}
		}
		\label{tbl:cxr_performances}
	\end{table}
	
	\begin{figure}[t!]
		\centering
		\IfFileExists{figures/samples_cxr/CXR_162_1946_108_PA-crop.pdf}{}{\immediate\write18{pdfcrop 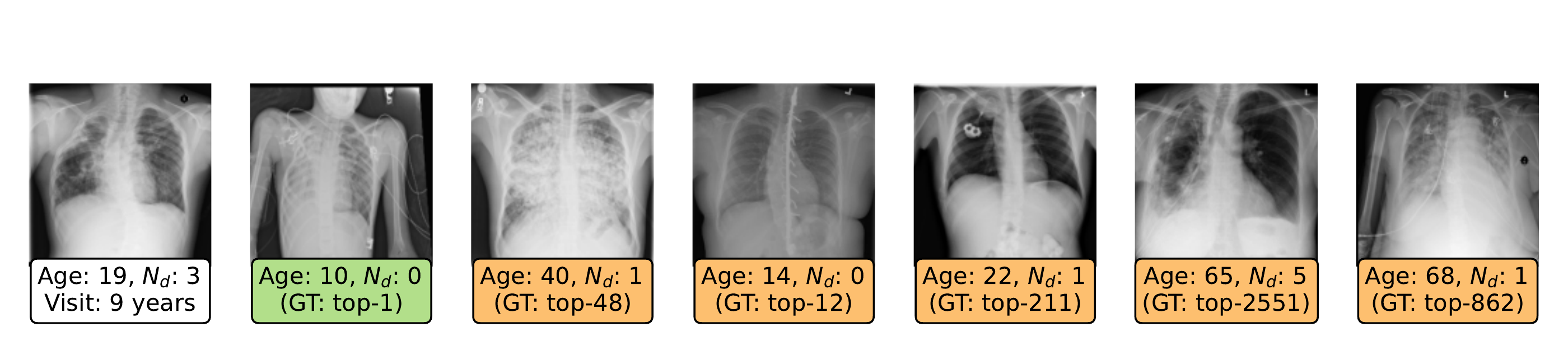}}
		\IfFileExists{figures/samples_oai/OAI_7_9399129_48_L-crop.pdf}{}{\immediate\write18{pdfcrop 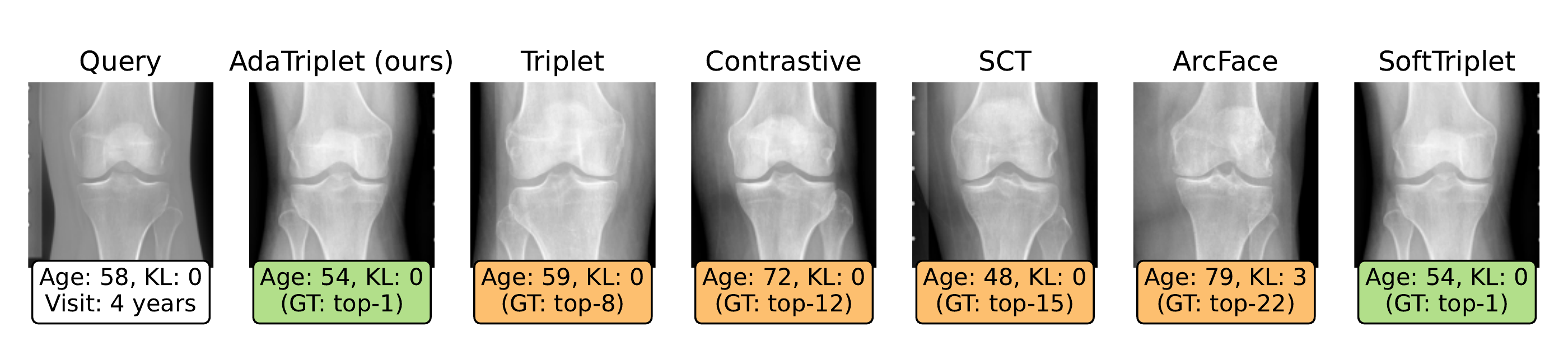}}
		\IfFileExists{figures/samples_oai/OAI_25_9547902_72_L-crop.pdf}{}{\immediate\write18{pdfcrop 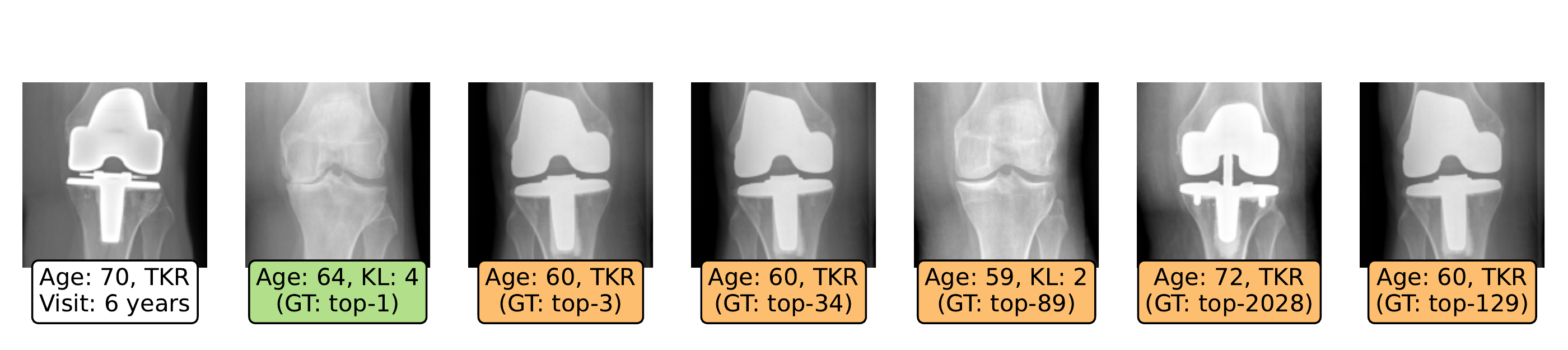}}
		\IfFileExists{figures/samples_oai/OAI_36_9402055_96_L-crop.pdf}{}{\immediate\write18{pdfcrop 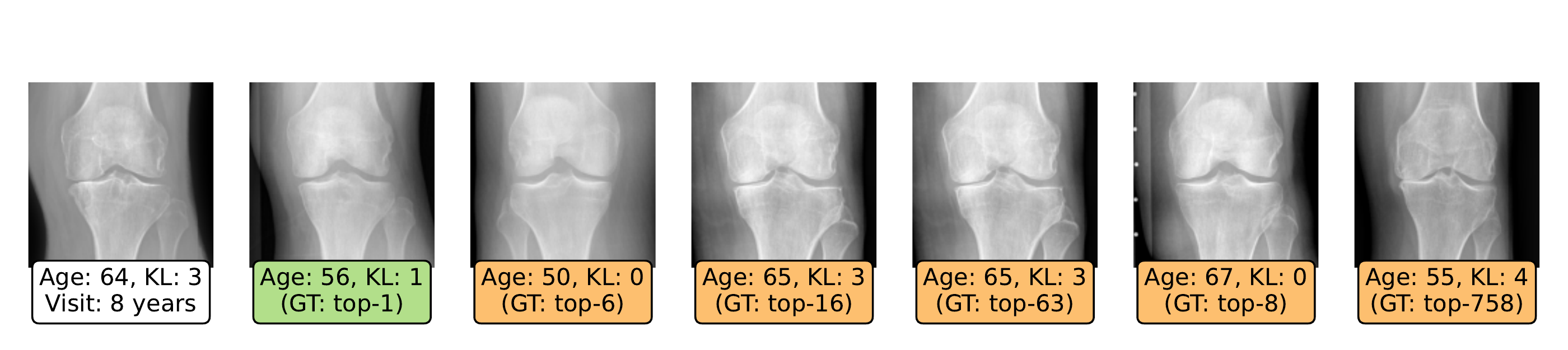}}
		\IfFileExists{figures/samples_oai/OAI_48_9461404_96_R-crop.pdf}{}{\immediate\write18{pdfcrop 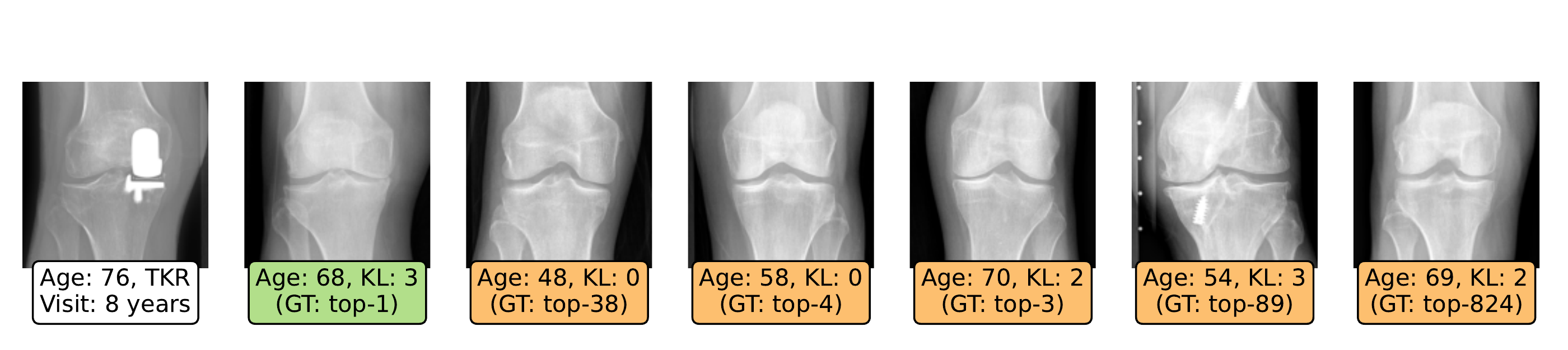}}
		\IfFileExists{figures/samples_oai/OAI_47_9378288_96_R-crop.pdf}{}{\immediate\write18{pdfcrop 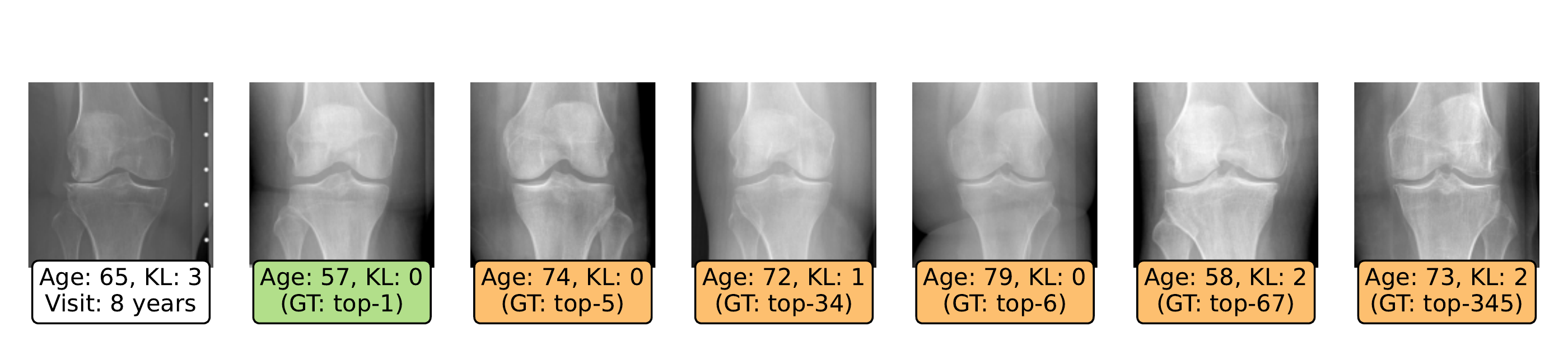}}
		
		\IfFileExists{figures/samples_cxr/CXR_162_1946_108_PA-crop.pdf}{}{\immediate\write18{pdfcrop figures/samples_cxr/CXR_162_1946_108_PA.pdf}}
		\IfFileExists{figures/samples_cxr/CXR_171_1836_84_AP-crop.pdf}{}{\immediate\write18{pdfcrop 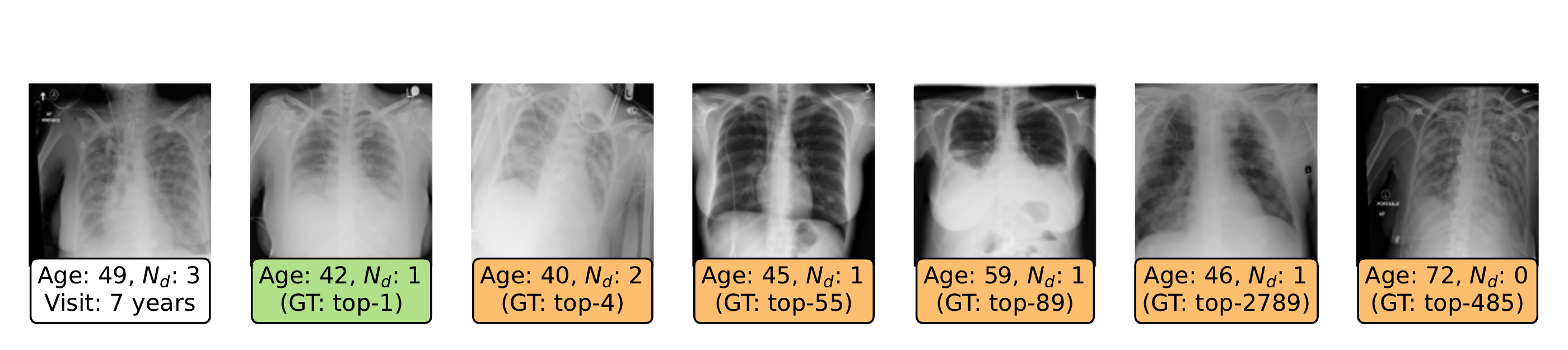}}
		\IfFileExists{figures/samples_cxr/CXR_195_4808_48_AP-crop.pdf}{}{\immediate\write18{pdfcrop 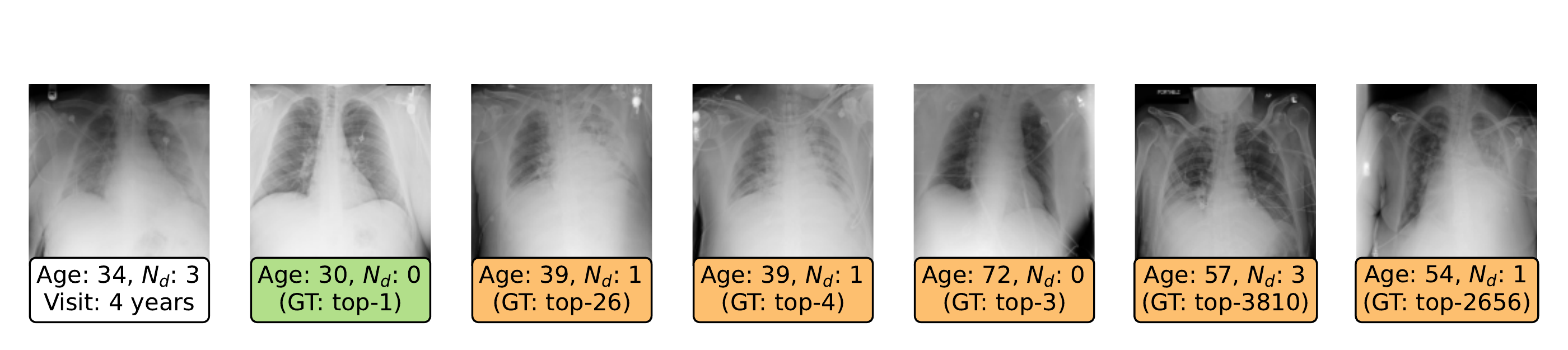}}
		\IfFileExists{figures/samples_cxr/CXR_83_23325_72_AP-crop.pdf}{}{\immediate\write18{pdfcrop 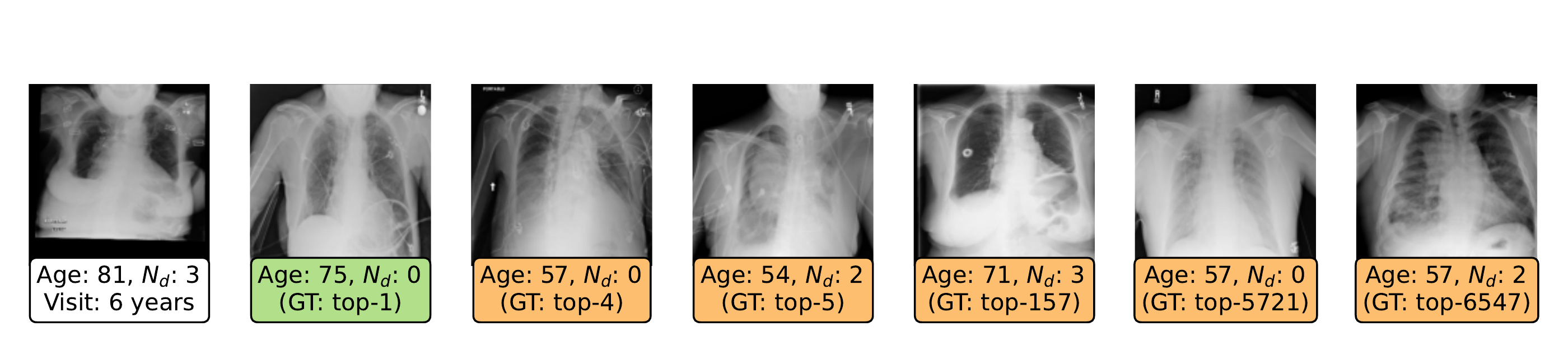}}
		
		\includegraphics[width=1\linewidth]{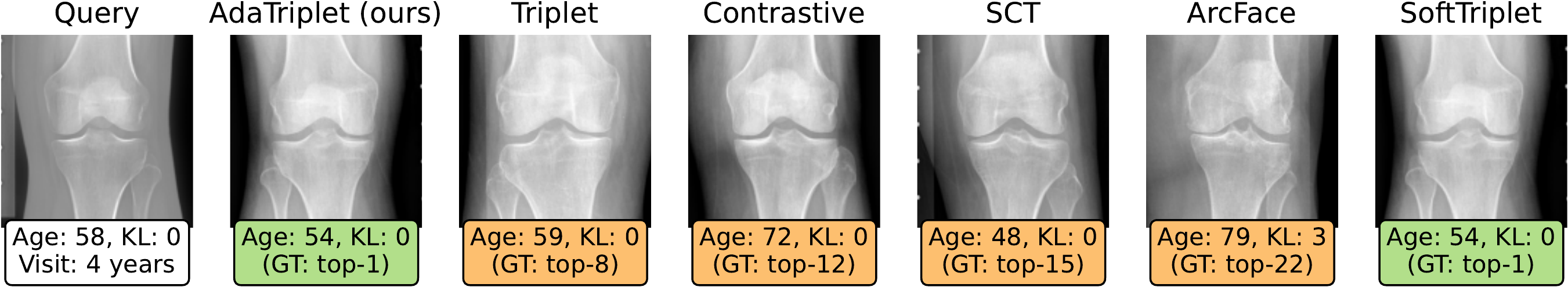} \\
		\includegraphics[width=1\linewidth]{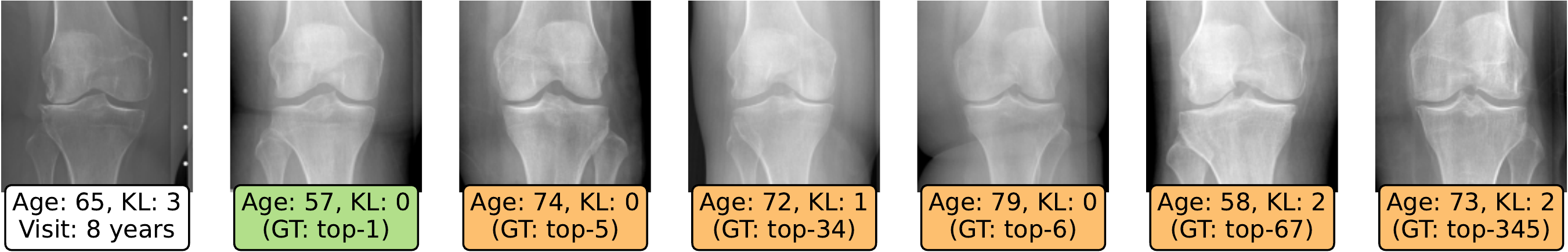} \\ 
		\includegraphics[width=1\linewidth]{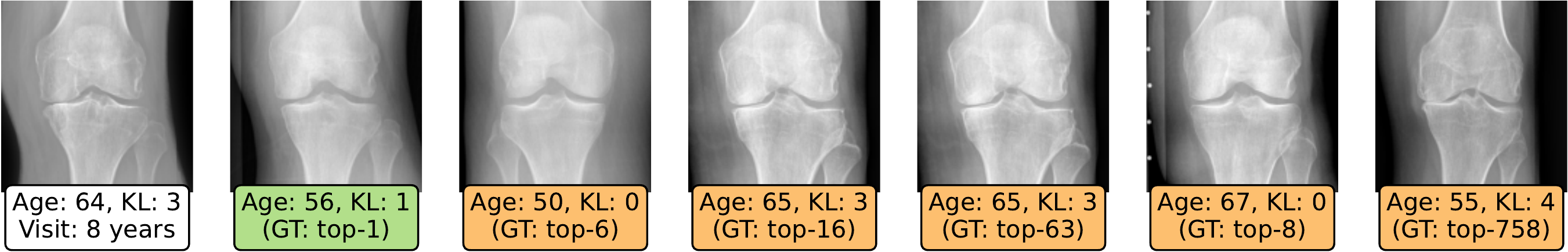} \\
		\includegraphics[width=1\linewidth]{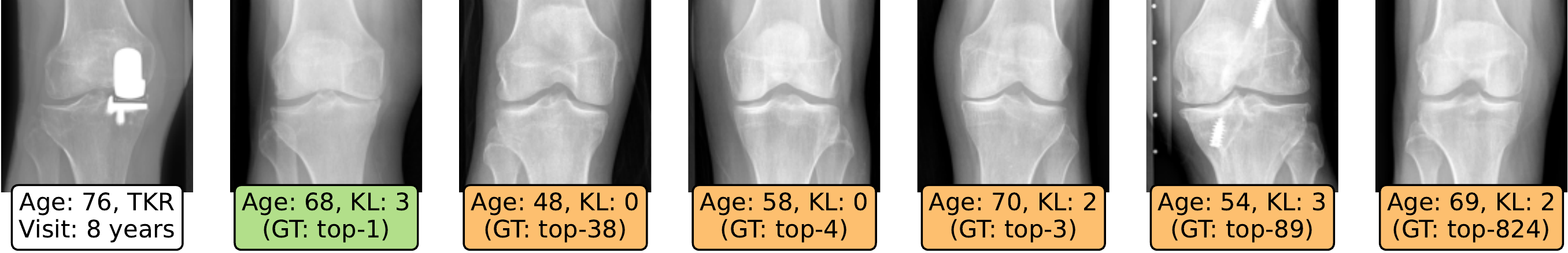} \\
		\includegraphics[width=1\linewidth]{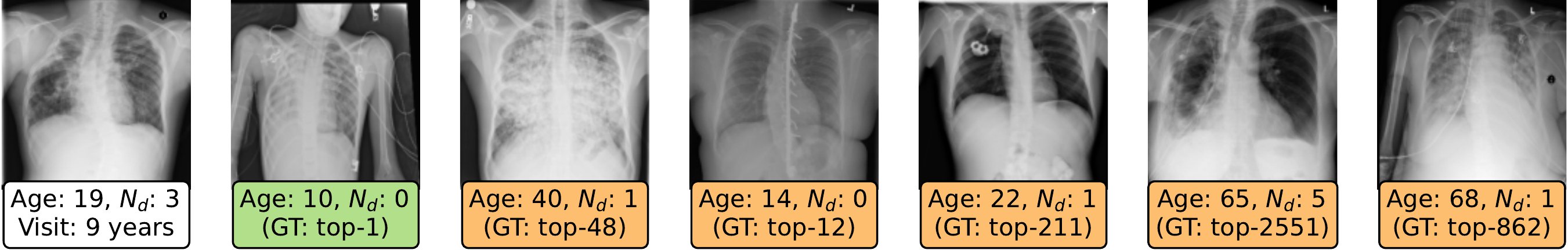} \\
		\includegraphics[width=1\linewidth]{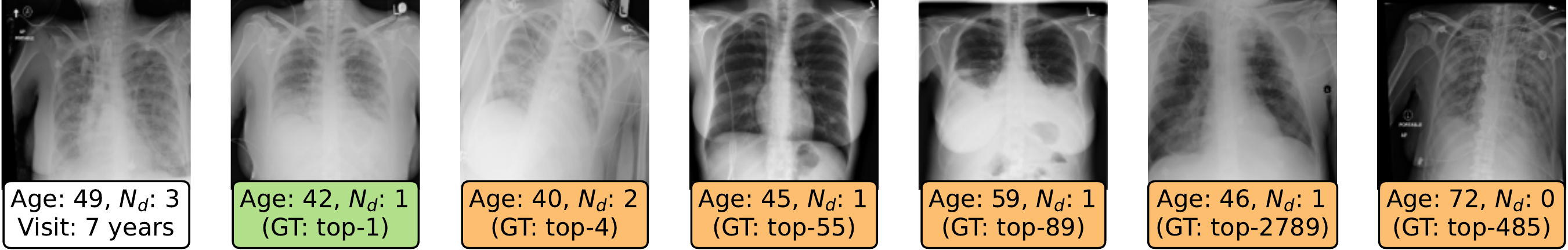} \\
		\includegraphics[width=1\linewidth]{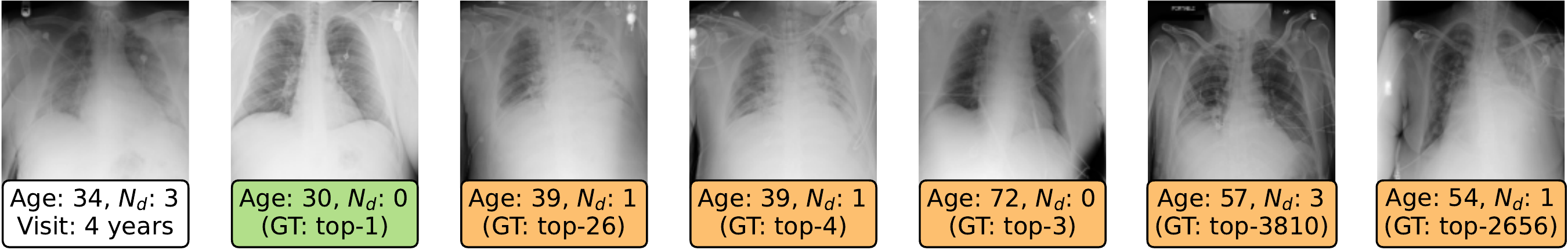} \\
		\includegraphics[width=1\linewidth]{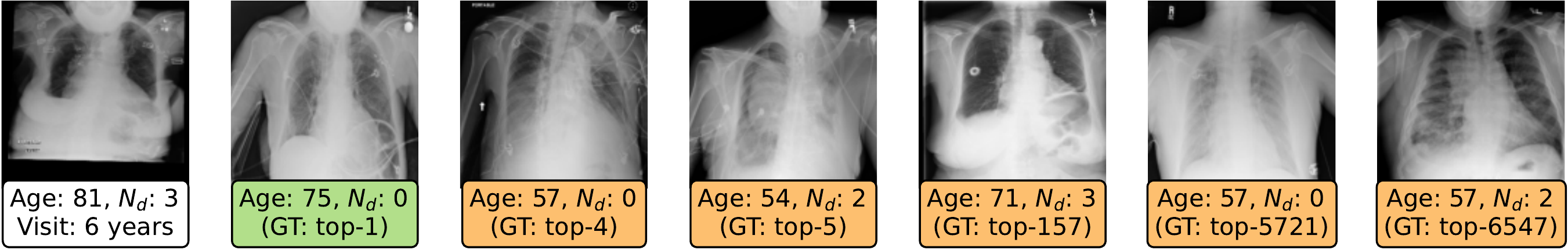}
		\caption{\small Matching samples of our method and the other baselines. Columns 2-7 are the top-$1$ matched images of the corresponding methods. Top-$k$ indicates the position of ground truth (GT). Green: top-1 prediction is the correct person (GT is top-$1$), orange: otherwise. KL means the Kellgence-Lawrence grade, assessing the stage of knee osteoarthritic severity. TKR indicates knees undergone total knee replacement surgery. $N_d$ is the number of thorax diseases.}
		\label{fig:matching_samples_more}
	\end{figure}
	
	
	
\end{document}